\documentstyle[aps,eqsecnum,pre,graphicx]{revtex}
\input epsf
\begin{document}
\draft

\title{Phase diagram of a model for $^3He-^4He$ mixtures in three dimensions}
\author{  A. Macio\l ek,$^1$  M. Krech,$^{2,3}$ and  S. Dietrich$^{2,3}$\\
 {\small $^1$\it   Institute of Physical Chemistry, 
             Polish Academy of Sciences,}\\
     {\small \it       Department III, Kasprzaka 44/52,
            PL-01-224 Warsaw, Poland,}\\
   {\small $^2$\it Max-Planck-Institut f{\"u}r Metallforschung,\\
Heisenbergstr. 3, D-70569 Stuttgart, Germany,\\}
{\small \it $^3$ Institut f{\"u}r Theoretische und Angewandte Physik, Universit{\"a}t Stuttgart,\\
Pfaffenwaldring 57, D-70569 Stuttgart, Germany\\}
    }
            
\date{\today}
\maketitle
\begin{abstract}
\baselineskip6mm
A lattice model of $^3$He -$^4$He mixtures which takes into account the
continuous rotational symmetry $O(2)$ of the superfluid
degrees of freedom of $^4$He is studied in the molecular-field approximation
and by Monte Carlo simulations in three dimensions. In contrast to its
two-dimensional version, for reasonable values of the interaction parameters
the resulting phase diagram resembles that observed
experimentally for $^3$He -$^4$He mixtures, for which phase
separation occurs as a consequence of the superfluid transition.
The corresponding continuum  Ginzburg-Landau model with two order parameters
 describing $^3$He-$^4$He
mixtures near tricriticality is derived from the considered lattice model.
All coupling constants appearing in the continuum model are explicitly
expressed in terms of the mean concentration of $^4$He, the temperature, 
and the microscopic interaction parameters characterizing the lattice system.

\end{abstract}

\section{Introduction}
\label{sec:intro}
Spatial confinements of systems undergoing continuous phase transitions perturb
the fluctuation spectrum of the corresponding ordering degrees of freedom.
This leads to a dependence of the free energy of the systems on the distance
between the confining walls which can be expressed in terms of universal scaling function.
 Their gradient renders the so-called critical Casimir forces which
are the analogues of the well known electro-magnetic Casimir or dispersion forces.

Recent developments in the theory of Casimir forces in critical
 and correlated fluids
~\cite{krech:99:0} have provided a strong motivation for testing them
experimentally in various systems.
Capacity studies of  $^4$He wetting films  near the 
superfluid transition $T_{\lambda}$~\cite{garcia:99:0}
have confirmed the existence  of the critical Casimir effect  and for temperatures
$T>T_{\lambda}$ quantitative agreement with  corresponding theoretical predictions has 
been found~\cite{krech:91:0,krech:92:0,krech:92:1}.
Similar effects have been observed for wetting films of 
 binary liquid mixtures near the critical
end-point of their demixing transition~\cite{law:99:0}.
Additional evidence for the critical Casimir force and detailed data  have
 been reported  for wetting  films at solid substrates of  $^3$He-$^4$He
mixtures near  their  tricritical point~\cite{garcia:02:0,balibar:02:0}. 
These latter experiments have also raised  new interesting challenges
for the theory, which have motivated the present work.
Among them is the sign and the amplitude  of the Casimir force
or, more generally, the form of its scaling function for different  values of the concentration of $^3$He
 atoms.
Theory predicts that these features of the Casimir force  crucially depend on
the type of boundary conditions which the confining surfaces impose on the order parameter~\cite{krech:99:0}.
For example, the force should be attractive for  symmetric boundary
conditions and repulsive for nonsymmetric boundary conditions.
The distinction between the surface universality classes  is also relevant.
In the case of pure $^4$He films near the $\lambda$-point the boundary conditions at the two confining  interfaces of the wetting layer seem to be very well
 approximated by symmetric  Dirichlet boundary conditions forming the so-called
ordinary surface universality class, because the quantum-mechanical wave function that describes
 the superfluid state vanishes at  both interfaces~\cite{krech:99:0,garcia:99:0}.
For films of  $^3$He-$^4$He mixture  the situation is less clear.
In  these systems  a $^4$He-rich layer
forms near the substrate-fluid interface, which may become superfluid
already above the $\lambda$-line~\cite{laheurte:77:0,leibler:84:0,macqueeney:84:0}, whereas
there is an enrichment of $^3$He near the opposing fluid-vapor interface.
Thus the two interfaces impose a nontrivial concentration profile, which in turn couples
to the superfluid order parameter.
The experiment of Ref.\cite{garcia:02:0} reports
 a repulsive Casimir  force at the tricritical point but
 it is not immediately obvious that the concentration profile  induces effectively 
nonsymmetric boundary conditions for the superfluid order parameter, i.e., symmetry-breaking boundary
 conditions  at the  substrate-fluid interface
 (also known as the so-called extraordinary or normal universality class) and   Dirichlet boundary conditions
 at the fluid-vapor interface.
The superfluid order parameter possesses  a continuous $O(2)$ symmetry so that
 if the layer
 is effectively two-dimensional  it is in the
Kosterlitz-Thouless phase
 with the superfluid order-parameter equal to zero~\cite{kosterlitz:80:0}.
 On the other hand, upon approaching the $\lambda$-line from the high temperature side
 the superfluid layer  thickens due to the increase  of the correlation length
and a dimensional crossover to a three-dimensional 
 superfluid phase with non-zero order-parameter should take place~\cite{dietrich:91:0}.
In order to be able to interpret and to
understand  the features of the Casimir force and other
surface and finite-size effects  in $^3$He-$^4$He mixture films near the
tricritical point
 systematic studies  of  a model system are needed.
Due to the universal character of the critical Casimir force
it is sufficient to choose as a model system  one which belongs to
the same universality class as the actual physical system. 
 The prerequisite of such future  studies is a detailed analysis 
 of the  bulk properties of such a suitable model.
This is the purpose of  the present paper.
The model should resemble the main features of the bulk phase diagram of 
$^3$He-$^4$He mixtures in three dimensions ($d=3$) and take  into 
 account the continuous rotational O(2) symmetry of the superfluid
degrees of freedom of $^4$He. 

The general features of phase separation and superfluidity  in 
three-dimensional mixtures  
of liquid  $^3$He and $^4$He are well known from experiments~\cite{graf:67:0}.
In pure $^4$He  there is a transition from a normal fluid to 
a superfluid phase characterized by a complex order parameter. 
If $^4$He is diluted with $^3$He, the superfluid transition
 temperature  is depressed.
Simultaneously, the tendency toward phase separation increases and at a 
critical  $^3$He concentration the mixture undergoes a
 first-order phase transition into a $^4$He-rich and a $^3$He-rich  phase,
 of which only the $^4$He-rich phase is superfluid. 
In the temperature - $^3$He concentration plane $(T,x)$ the line 
of second-order superfluid transitions $T_s(x)$ meets the
 boundary of the two-phase coexistence region 
at the tricritical point $(T_t\approx 87mK,x_t\approx 0.67)$; $x =
N_3 / (N_3 + N_4)$, where $N_i$, $i = 3, 4$, denotes the number of atoms of
$^3$He or $^4$He, respectively.

In this paper we consider a simple lattice model known in the literature 
 as  the Vectoralized Blume-Emery-Griffiths model
 (VBEG)~\cite{cardy:79:0,berker:79:0}.
It  is defined in Sec.~\ref{sec:mod}. The bulk phase diagram of the VBEG model 
 was investigated only in spatial  dimensions $d=2$
  by means of Migdal-Kadanoff
recursion relations~\cite{cardy:79:0,berker:79:0}; no tricritical point was found for any value of the  model parameters.
Here we study the  three-dimensional version of this model within
 the molecular-field approximation
and by Monte Carlo simulations,  and we demonstrate that for
 reasonable values of interaction parameters the resulting phase
 diagram resembles topologically
 that observed experimentally for  three dimensional
mixtures, for which the phase separation appears as a consequence of the
 superfluid ordering.
 This is carried out in Sec.~\ref{sec:mf} and
Sec.~\ref{sec:MC}, respectively.
In Sec.~\ref{sec:LG} we  derive a two-parameter continuous
 Landau-Ginzburg (GL)
 model describing bulk $^3$He-$^4$He mixtures near
a tricritical point starting from the modified VBEG model.
The GL  approach has many advantages and it is worthwhile to have a LG model
with coupling constants explicitely related to the measured quantities, such like like temperature, superfluid density, concentration of $^3$He atoms, or parameters describing interactions.
We close our paper with a summary and conclusions.

\section{The model.}
\label{sec:mod}
We consider a simple lattice  model of   $^3$He-$^4$He mixtures
which takes into account the continuous rotational O(2)
 symmetry of the superfluid
degrees of freedom of $^4$He. It is a  vectorial generalization of the
 spin-1 model 
used by Blume, Emery, and Griffiths (BEG) ~\cite{blume:71:0}
  for describing the
$\lambda$-line and the tricritical point in $^3$He-$^4$He mixtures.
In this model each simple cubic lattice  site $i$ 
is associated with an occupation  variable  $t_i$, taking the values 0 or 1,
and a phase $\theta _i$ $( 0\le \theta _i< 2\pi)$ which mimics
 the phase of the $^4$He  wave function.
A $^3$He atom at site $i$ corresponds  to $t_i=0$ and a $^4$He atom 
to $t_i=1$. Since the model in this reduced version does not allow
for unoccupied sites, the model does not exhibit a vapor phase.
(In future studies the model can be generalized to incorporate
the vapor phase; here it is left out for reasons of simplicity.)
 $\theta _i$ reflects the superfluid degrees of freedom.
The model Hamiltonian consists of a lattice gas part describing
 a binary mixture   and
a term responsible for
  the ``superfluid'' ordering.
 Since only $^4$He atoms couple to the superfluid order parameter
 the Hamiltonian is taken as
\begin{equation}
\label{eq:ham}
{ \cal H}=-J\sum _{<ij>}t_it_j\cos (\theta _i-\theta _j) - K\sum _{<ij>}t_it_j+\Delta\sum _it_i
\end{equation}
where    the first two sums  are over  nearest-neighbor pairs
$<ij>$,  and  the last sum is over all lattice sites. 
The lattice constant $a$ is taken to be equal to 1.

In the lattice gas model of the  $^3$He-$^4$He binary mixture  the 
coupling constant $ K$ and the field $\Delta$ are related to the effective
$^{\alpha}$He-$^{\beta}$He interactions 
  -$K_{\alpha \beta}$~\cite{bell:89:0}
\begin{equation}
\label{eq:parK}
K=K_{33}+K_{44}-2K_{34},
\end{equation}
and to the chemical potentials $\mu _3$ and $\mu _4$  of  $^3$He and $^4$He,
 repectively,
\begin{equation}
\label{eq:pardelta}
\Delta=\mu _3-\mu _4+2z(K_{33}-K_{34}),
\end{equation}
where  $z$ is  the coordination number
 of the lattice ($z=2d$, where $d$ is  the 
spatial dimension of the system; $z=6$ in the present case).
 
In the liquid the effective interactions 
 $K_{\alpha \beta}$ are different for different $\alpha$
 and $\beta$ due to the differences of mass and statistics  between $^3$He
 and $^4$He atoms.
The coupling constant $J$ is related~\cite{berker:79:0,book}  to a bare, 
 superfluid  density $\rho _0(T)$ by
\begin{equation}
\label{eq:paJ}
J=\hbar ^2\rho _0(T)a^{(d-2)}/m^2
\end{equation}
where $m$ is the mass of  a $^4$He atom. $a$ is the mean interparticle spacing
(the lattice parameter in the lattice model).
The superfluid density can be measured from the velocity of third
 sound and from 
the response of a torsional oscillator \cite{bishop:78:0};
 it has units of mass per unit volume (or area in two dimensions).
Here we are concerned only with the case  $J>0$ and $K>0$.

When all occupation numbers $t_i$ are equal to 1,  up to constants
 the first term in Eq.(\ref{eq:ham})
corresponds to  the classical $XY$ model 
(the planar rotator model) for pure $^4$He.
Therefore, in the limit of $\Delta\to -\infty$ the partition function of the model  reduces to that of $XY$  model up to a factor $e^{KzN}$ where $N$ is the
number of lattice sites.

The model as defined above is known in the literature 
 as  the Vectoralized Blume-Emery-Griffiths model (VBEG).
It was first proposed by Berker and Nelson~\cite{berker:79:0} and, independently, by
Cardy and Scalpino~\cite{cardy:79:0} 
 to describe thin
{\it films} of  $^3$He-$^4$He mixtures, for which  the mechanism of the 
superfluid transition is different from that in $d=3$;
there is no spontaneous  breaking of the continuous symmetry in $d=2$
~\cite{mermin:66:0}, i.e., the order parameter does not become nonzero below
 the transition temperature.
The superfluid transition in  films  of  $^3$He-$^4$He mixtures
 is  of the Kosterlitz-Thouless type~\cite{kosterlitz:80:0}.

In $d=2$  the phase diagram of  the VBEG model was
obtained by means of 
the Migdal-Kadanoff  renormalization-group method
~\cite{cardy:79:0,berker:79:0}.  
Its features are qualitatively similar to those  observed experimentally
for the corresponding three-dimensional mixtures, except that there is
 no true tricritical point for any value of the model parameters.
The line of the superfluid transitions  ($\lambda$-line) is connected to the
phase-separation curve  by a critical endpoint at a temperature
 distinctly lower than the phase-separation critical temperature.
 Thus upon lowering the temperature 
the system first phase separates  into two normal fluids with different
 concentrations of $^3$He. At lower temperatures, there is  phase separation
 into a superfluid phase with a  low $^3$He concentration and into a normal
 fluid with a  high $^3$He concentration. 

In this paper we are interested in the corresponding three-dimensional systems.
We determine the phase diagram of the VBEG model 
within  mean-field approximation and by Monte Carlo simulations.

\section{Molecular Field Approximation}
\label{sec:mf}
\subsection{ Free energy}
\label{sub:1}
In this section we determine the phase diagram of the VBEG model whithin the molecular-field approximation.
It is derived from the variational method  based upon approximating the 
total equilibrium  density matrix by a product  of local site density matrices
$\rho _i$~\cite{book}.

The variation theorem for the free energy reads:
\begin{equation}
\label{eq:var}
F\le F_{\rho}=Tr(\rho{\cal H})+(1/\beta)Tr(\rho \ln\rho)
\end{equation}
where $F$ is the exact free energy and $F_{\rho}$ is an approximate free energy
 associated with the density matrix $\rho$; $\beta =1/k_BT$.
The minimum of $F_{\rho}$  with respect to the variation of $\rho$ subject to 
the constraint $Tr\rho =1$ is attained  for the equilibrium density  matrix,
$\rho=e^{-\beta {\cal H}}/Tr(e^{-\beta {\cal H}})$.

Whithin mean-field theory the density matrix is approximated by
\begin{equation}
\label{eq:denmat}
\rho=\rho_0=\prod_{i=1}^N\rho_i
\end{equation}
where in homogeneous bulk systems the local density matrix  $\rho _i$ is independent of the site $i$.
For the Hamiltonian given by Eq.(\ref{eq:ham})
the variational mean-field free energy per site is
\begin{eqnarray}
\label{eq:varfe}
\frac{F_{\rho _0}}{N}\equiv \frac{\Phi}{N} &= &-\frac{{\tilde K}}{2}(Tr(t_i\rho_i))^2-\frac{{\tilde J}}{2}\left[(Tr(t_i\cos\theta _i\rho _i))^2+(Tr(t_i\sin\theta _i\rho _i))^2\right]  \nonumber \\
&+&\Delta Tr(t_i\rho _i)+(1/\beta)Tr(\rho _i\ln\rho_i)
\end{eqnarray}
where  ${\tilde K}=zK$ and ${\tilde J}=zJ$.
To determine variational minima to Eq.(\ref{eq:varfe})
we treat the local site  density $\rho _i$ as a variational  function,
and the  best functional form in terms of $t_i$ and $\theta _i$ is 
obtain by minimizing  $ F_{\rho _0}$ with respect to $\rho _i$.
In this procedure, however, the connection between  $ F_{\rho _0}$
and the  Hemholtz free energy functional of the order parameters
is not straightforward~\cite{comment}.
Minimizing 
$\Phi/N+\eta Tr(\rho _i)$  
with respect  to  $\rho_i$ and with  $\eta $ as a Lagrange multiplier one obtains
\begin{equation}
\label{eq:solden}
\rho _i=e^{-\beta h_i}/Tr(e^{-\beta h_i})
\end{equation}
where  $h_i$ is the single-site  molecular Hamiltonian  given by
\begin{equation}
\label{eq:h}
 h_i=-{\tilde K}(Tr(t_i\rho _i))t_i-{\tilde J}
\left[(Tr(t_i\cos\theta _i))t_i\cos\theta _i+(Tr(t_i\sin \theta _i))t_i\sin \theta _i\right]+\Delta t_i.
\end{equation}

We  define the following order parameters:
\begin{equation}
\label{eq:q}
Q\equiv 1-x=<t_i>
\end{equation}
and 
\begin{equation}
\label{eq:m}
M_x= <t_i\cos \theta _i>\qquad M_y= <t_i\sin \theta _i>.
\end{equation}
$Q$ corresponds to the concentration of  $^4$He, $x$ to the concentration of 
$^3$He, and $M_x, M_y$
are the components of the two-dimensional superfluid order parameter
 ${\bf M}=(M_1,M_2)$
with $M={{\sqrt{ M_x^2+M_y^2}}}$.
Within this approximation  $Q(\Delta,T)$ and $M(\Delta,T)$ are given by 
two coupled self-consistent equations:
\begin{equation}
\label{eq:eqQ}
Q=\frac{I_0(\beta{\tilde J}M)}{e^{\beta(-{\tilde K}Q+\Delta)}+I_0(\beta{\tilde J}M)}
\end{equation}
and
\begin{equation}
\label{eq:eqM}
M=\frac{I_1(\beta{\tilde J}M)}{e^{\beta (-{\tilde K}Q+\Delta)}+I_0(\beta {\tilde J}M)}
\end{equation}
where $I_0(z)$ and $I_1(z)$ are  modified Bessel functions~\cite{abramowitz}.

The equilibrium free energy  $\Phi (\Delta,T)$ is given by
\begin{equation}
\label{eq:freeeq}
\Phi (\Delta,T)/N=\frac{{\tilde K}}{2}((Q(\Delta ,T))^2+\frac{{\tilde J}}{2}(M(\Delta,T))^2+(1/\beta )\ln(1-Q(\Delta,T)).
\end{equation}
Most parts of the phase diagram  can only be determined by solving
 the equations for $Q$ and $M$ 
numerically. Some regions, however, can be studied analytically.

\subsection{ $\lambda$-line and tricritical point}
\label{subsec:2}

In order to find the line of critical points on which  second-order
 transitions from the 
normal ($M=0$) to the the superfluid ($M\ne 0$) state take place,
one needs the  thermodynamic potential in terms of  the order parameter $M$,
\begin{equation}
\label{eq:thermA}
A(M,\Delta, T)=\Phi-MH,
\end{equation}
where   $H$ is a  field conjugate to $M$: 
\begin{equation}
\label{eq:magf}
H=-\frac{\partial A}{\partial M}.
\end{equation}
The conditions for the critical points are:
\begin{equation}
\label{eq:cond}
\frac{\partial H}{\partial M}=\frac{\partial^2 H}{\partial M^2}=0,
\qquad  \frac{\partial^3 H}{\partial M^3}>0,
\end{equation}
and  the tricritical point is determined by
\begin{equation}
\label{eq:condtri}
\frac{\partial H}{\partial M}=\frac{\partial ^2H}{\partial M^2}=\frac{\partial^3 H}{\partial M^3}=\frac{\partial^4 H}{\partial M^4}=0,
\qquad \frac{\partial^5 H}{\partial M^5}>0.
\end{equation}
To find $H$ as a function of $M$ we use
the analogue of Eq.(\ref{eq:eqM}) for $H\ne 0$:
\begin{equation}
\label{eq:eqMH}
M=\frac{I_1({\beta\tilde J}M+\beta H)}{e^{\beta(-{\tilde K}Q+\Delta)}+I_0(\beta{\tilde J}M+\beta H)}.
\end{equation}
Since this equation  cannot be inverted explicitly,
 we expand it around $H=0$  keeping only  terms linear in $H$, and find
\begin{equation}
\label{eq:eqH_M}
\beta H=\frac{I_1(\beta{\tilde J}M)-MI_0(\beta{\tilde J}M)+Me^{\beta(-{\tilde K}Q+\Delta)}}
{MI_1(\beta{\tilde J}M)-(1/2)(I_0(\beta{\tilde J}M)-I_2(\beta{\tilde J}M))}.
\end{equation}
Applying the conditions formulated in Eq.(\ref{eq:cond})-(\ref{eq:eqH_M}) yields
 the whole line of critical points, i.e., the $\lambda$-line $T_s(x)$:
\begin{equation}
\label{eq:critcurve}
T_s(x)=\frac{{\tilde J}(1-x)}{2}=\frac{{\tilde J}Q}{2}.
\end{equation}
It follows that, as the  concentration of  $^3$He increases 
from zero, $T_s$ decreases linearly from $T_s(0)={\tilde J}/2$.
The critical curve $\Delta = \Delta _s(T)$ in the $(\Delta, T)$ plane 
can be obtained by first solving Eq.(\ref{eq:eqQ}) for $\Delta$ (here and in the following
we include $k_B$ into $T$)
which gives
\begin{equation}
\label{eq:eqDelta}
\Delta (Q,T)=T\ln (1-Q)-T\ln Q+{\tilde K}Q+T\ln I_0({\tilde J}M),
\end{equation}
 and then evaluate Eq.(\ref{eq:eqDelta})
for $M=0$ and $Q=2T/{\tilde J}$ (see Eqs.(\ref{eq:q}) and (\ref{eq:critcurve})).

The line of  second-order phase transitions  ends  at the tricritical point
$(T_t,x_t)$ where the transition changes to a first-order one.
From Eqs. (\ref{eq:condtri}) and (\ref{eq:eqH_M})
one obtains for  the temperature $T_t$ 
\begin{equation}
\label{eq:tritemp}
T_t/ T_s(0)=\frac{1+2K/J}{2+2K/J}  
\end{equation}
and for the concentration $x_t$ 
\begin{equation}
\label{eq:tritempQ}
 T_t/ T_s(0)=1-x_t,
\end{equation}
provided $ \partial^5 H/\partial M^5>0$ holds for the chosen
value of $K/J$. It is possible that not all critical points on this so-called critical line
represent equilibrium phase transitions because the latter ones are preempted by
first-order demixing transitions. Thus it can be that only a portion of this  so-called critical
line gives the $\lambda$-line, the rest being metastable.

\subsection{ Demixing}
\label{subsec:3}

For the disordered  phase with $M=0$
 one  can easily find the
first-order phase separation  line from the $^3$He-rich 'normal' fluid to
the $^4$He-rich 'normal' fluid.
The phase separation  is associated with an instability loop including a range
of $Q$ values for which  $\partial \Delta /\partial Q>0$ and the critical point is given
 by  $\partial \Delta /\partial Q=\partial ^2 \Delta /\partial Q^2=0$.
These last two relations together with Eq.(\ref{eq:eqDelta}), evaluated at $M=0$,
are satisfied  if  $Q_c=1/2$ and $T_c={\tilde K}/4$. 
The critical value $\Delta _c$ of $\Delta$ is ${\tilde K}/2$.
Inserting  $\Delta =\Delta _c$ and  $M=0$ into
 Eq.(\ref{eq:eqDelta}) gives
\begin{equation}
\label{eq:coexcurve}
T\ln\frac{1-Q}{Q}-(1/2){\tilde K}(2Q-1)=0
\end{equation}
For $T<{\tilde K}/4$ this equation  has pairs of solution
$(Q,1-Q)$. For $M=0$, i.e., above the critical line  in the $(Q,T)$
plane, these solution  form
the coexistence curve which is symmetric  about $Q=1/2$ or $x=0.5$.
 For temperatures lower than the intersection 
temperature  $T_I$ of  the critical line  with the curve given by Eq.(\ref{eq:coexcurve})
the phase rich in $^4$He becomes superfluid and Eq.(\ref{eq:coexcurve}) no
longer represents the coexistence curve.

In order to find what types of  phase diagrams the present model provides
 we look for the phase-separation instability  on the critical curve as determined in Subsec.~\ref{subsec:2}
and how it is located with respect to the intersection point $P_I=(Q_I,T_I)$.
Depending on the ratio $K/J$ there are three
 possibilities which give three different types
of the phase diagram: (i) the instability point $P_t$ lies below the
 intersection point $P_I$, (ii) $P_t$ lies above $P_I$, and (iii) the
 critical point of the transition between   $^3$He- and $^4$He-rich 'normal'
 fluids falls into the instability  range initiated at  $P_t$.

A sufficient condition for an instability loop leading to phase separation 
is $\partial \Delta/\partial Q >0$.
Using  Eq.(\ref{eq:eqDelta}) and the relation $Q=MI_0(\beta{\tilde J}M)/I_1(\beta{\tilde J}M)$
one finds
\begin{equation}
\label{eq:b1'}
\left(\frac{\partial \Delta}{\partial Q}\right)_{Q=Q^*+}=\frac{-{\tilde J}}{2(1-Q^*)}+{\tilde K}+{\tilde J} 
\end{equation}
whereas 
\begin{equation}
\label{eq:b1}
\left(\frac{\partial \Delta}{\partial Q}\right)_{Q=Q^*-}=\frac{-{\tilde J}}{2(1-Q^*)}+{\tilde K},
\end{equation}
where $Q^*$ is the critical value of $Q$ for superfluid ordering given by
Eq. (\ref{eq:critcurve}).
Thus  $\left(\partial \Delta/\partial Q\right)_{Q=Q^*-}> \left(\partial \Delta /\partial Q\right)_{Q=Q^*+}$
and the instability will occur on the ordered side of the critical curve when
 $\left(\partial \Delta /\partial Q\right)_{Q=Q^*+}=0$.
From Eq. (\ref{eq:b1'}) the coordinates of the instability point $P_t$ are given by
Eq. (\ref{eq:tritemp}), i.e., they are exactly the same as as those of the tricritical point.
We find numerically that $P_I$ and $P_t=(Q_t,T_t)$ coincide for $K/J\approx 2.01681$.

For $K/J> 2.01681$  case (i) is realized, i.e., the instability point $P_t$
 lies  inside the coexistence curve between  the $^3$He-rich 'normal' fluid and
the $^4$He-rich 'normal' fluid. 
 We have obtained numerically the phase diagram for $K/J=2.4$.  
It is shown in Figs.~\ref{fig:type1}(a)
 and \ref{fig:type1}(b) in the
$(\Delta,T)$ and $(x,T)$  plane, respectively. 
The dashed  line   in the $(\Delta,T)$ plane represents the
line of equilibrium  second-order phase transitions on the critical curve and thus represents
the $\lambda$-line.
This line terminates at the phase separation curve (solid  line) at the 
{\it critical end point E}.
CE is the line of  the first-order phase transitions between the $^3$He-rich  and the 
$^4$He-rich 'normal' fluids
with the critical point C. At the point E the curve CE turns into the line of first-order 
phase transitions between the $^3$He-rich 'normal' fluid and the $^4$He-rich superfluid.
The phase boundary $\Delta (T)$ between the $^3$He-rich  and the 
$^4$He-rich 'normal' fluid or the $^4$He-rich superfluid (represented by a 
solid  line)
 is expected to exhibit a singular
 curvature  $\sim |T-T_E|^{-\alpha}$ as $T$ approaches  the end point 
temperature from above or below~\cite{fisher:91:0}. $\alpha$ is the critical exponent 
describing the specific heat singularity on the critical line below the point E.
Since in mean-field theory $\alpha=0$,  there is 
no nonanalyticity at the end point whithin the present approach.
In the $(x,T)$ plane (see Fig.~\ref{fig:type1}(b))
the coexistence curve is smooth at $T=T_E$ on the $^3$He-rich side.

For $K/J< 2.01681$ the instability point $P_t$ lies outside the coexistence curve
for the  $^3$He-rich 'normal' fluid and the $^4$He-rich 'normal' fluid.
 Therefore, as $T$ 
decreases below $T_t$, the phase separation between  the   $^3$He-rich 'normal' fluid
and the $^4$He-rich superfluid commences at the point $P_t$
on the critical curve; hence $P_t\equiv A$ is a {\it tricritical point}.
The phase diagram  for $K/J=1.8$ is shown in Fig.~\ref{fig:type2}.
In the  $(\Delta,T)$ plane (Fig.~\ref{fig:type2}(a)) the first-order transition line
between the  $^3$He-rich 'normal' fluid and the $^4$He-rich superfluid 
which starts at A terminates at a triple point $D$ where it meets the first-order
 transition lines between the
 $^3$He-rich 'normal' fluid and the $^4$He-rich 'normal' fluid (curve CD)
 and between the $^3$He-rich 'normal' fluid and the $^4$He-rich superfluid.

For even smaller value of the ratio $K/J$ there are no longer two
distinguishable disordered phases, i.e., the line DC in Fig.~\ref{fig:type2}(a)
has shrunk to zero.
 The  critical point for coexistence between the  $^3$He-rich 'normal' fluid
and the $^4$He-rich 'normal' fluid, 
which occurs at $x_c=1/2$ and $T_c={\tilde K}/4$, disappears.
In Fig.~\ref{fig:type3} we present the phase diagram for $K/J=1$.
In the  $(\Delta,T)$ plane it exhibits a very simple form (Fig.~\ref{fig:type3}(a)).
 The $\lambda$-line meets   the first-order transition line between the
 $^3$He-rich 'normal' fluid and the $^4$He-rich superfluid 
  at the {\it tricritical point A}. The lines meet with a common tangent, a feature characteristic
of the mean-field approximation.
In the $(x,T)$ plane (Fig.~\ref{fig:type3}(b)) at the tricritical point 
the critical line $T_s(x)$
  has the same slope as  the 
 phase-separation curve on the $^3$He-rich side. 
The emergence  of the type of phase diagram shown in Fig.~\ref{fig:type3}
from  the one shown in  Fig.~\ref{fig:type2} takes place  at that value of $K/J$ for which there is an equilibrium between the phase at the critical point 
 C  and an ordered phase yielding $K/J\approx 1.4298$.
For $K/J$ slightly less than this value  $K/J=1.4298$,
  the tricritical
 point is located at $T_t\approx 0.397$ and $x_t\approx 0.206$.
Whithin mean-field theory  the $^3$He-rich side of the coexistence curve has a plateau for $0.8\lesssim x \lesssim 0.3$, i.e., right below the tricritical point  small changes in temperature lead to pronounced 
changes in the  concentration of $^3$He. 
As $K/J$ is reduced further the tricritical point shifts to larger values of
$x$  and smaller values of $T$. Also the shape of the coexistence curve changes; the plateau disappears and the concentration of $^3$He inceases  more uniformly with the temperature.

A phase diagram like that of Fig.~\ref{fig:type3} resembles qualitatively the experimental one~\cite{graf:67:0}, for which one finds for the
tricritical point 
$T_A/T_s(0)=0.4$ and $x_A=0.669$. In our model $x_A$ is, however,  always
smaller than  0.5.

\section{ Monte Carlo Simulations}
\label{sec:MC}
For the Monte Carlo treatment of the model Hamiltonian
 given by Eq.(\ref{eq:ham})
a $^4$He atom is represented by the normalized spin vector
\begin{equation}
\label{spin}
{\bf S}_i \equiv (\cos \theta_i, \sin \theta_i)
\end{equation}
for each lattice site $i$ in the spirit of the standard XY model. A $^3$He
atom on lattice site $i$ is represented as ${\bf S}_i \equiv (0,0)$.
Consequently the occupation number $t_i$ on lattice site $i$ is given by
$t_i = |{\bf S}_i|$ and the interaction $\cos(\theta_i-\theta_j)$ between two
$^4$He atoms is given by the computationally more favorable scalar product
$\cos(\theta_i-\theta_j) = {\bf S}_i \cdot {\bf S}_j$. The lattice is 
simple cubic with periodic boundary conditions and $L$ lattice sites
in each direction. The Monte Carlo algorithm is based on various standard
procedures which we discuss briefly in the following.

In order to explore the phase space of the model, two types of updates
are needed: (i) spin flip updates and (ii) particle insertion and deletion
updates. The spin flip updates are responsible for the creation of long -
ranged magnetic order which in our model represents the normal - superfluid
transition. This can be of first or second order depending on the
concentration $x=1-<t_i>$ of $^3$He ($t_i = 0$) in the system (see
Fig. \ref{fig:type3}). The particle insertion and deletion updates are
responsible for the demixing transition (phase separation) in our model.
This transition can also be first or second order depending on the coupling
constants in the model (see Figs. \ref{fig:type1} and \ref{fig:type2}).
In our simulation we are primarily interested in that regime of coupling
constants, for which the phase diagram resembles that of actual $^3$He -$^4$He
mixtures (Fig. \ref{fig:type3}) and therefore the possibility of a second
order (critical) demixing
transition is not taken into account for the selection of the Monte Carlo
moves. We therefore use the following methods in our Monte Carlo simulation:
(i) single particle insertion and deletion and (ii) single spin flip according
to the Metropolis algorithm \cite{Metropolis}, (iii) single cluster spin flip
according to the Wolff algorithm \cite{Wolff89}, and (iv) overrelaxation
updates of the spin degrees of freedom at constant configurational energy
\cite{CFL93}. For each particle insertion or single spin flip move the new
spin state is randomly selected from the even distribution on the unit circle.
The projection vector for the embedding part of the Wolff algorithm
\cite{Wolff89} is also chosen randomly from the even distribution on the unit
circle.

The above update methods are performed in sweeps over the whole lattice,
where {\em each} spin flip sweep (ii), (iii), or (iv) is preceded by a
Metropolis particle insertion and deletion sweep (i). Cluster updates of
the particle configuration according to the embedding algorithm of
Ref.\cite{Wolff89} are disregarded, because the critical demixing transition
will not be explored here.

The three basic Monte Carlo updates (ii) - (iv) outlined above are combined
according to the hybrid Monte Carlo idea \cite{Hybrid} to ensure efficient
configuration space exploration also for second order (critical) transitions
to long-ranged magnetic order. One hybrid Monte Carlo step consists of 10
individual steps each of which can be one of the updates listed above.
The Metropolis and the Wolff algorithm work the standard way, in which  the
acceptance probability $p$ of a proposed spin flip in the Metropolis
algorithm is defined by the local heat bath rule
\begin{equation} \label{heatbath}
p(\Delta E) = 1/[\exp(\Delta E / k_B T) + 1],
\end{equation}
where $\Delta E$ is the change in configurational energy of the proposed move.
The overrelaxation part of the algorithm performs a microcanonical update of
the configuration by sequentially reflecting each spin in the lattice at
the direction of the local field, i.e., the sum of the nearest neighbor
spins, such that its contribution to the energy of the whole configuration
remains constant. The implementation of this update scheme is straightforward,
because according to Eq.(\ref{eq:ham}) the energy of a spin with respect to its
neighborhood has the functional form of a scalar product. The form hybrid Monte
Carlo step depends on the region of the phase diagram to be explored. In
the vicinity of the first-order phase separation line typically six
Metropolis (M), one single cluster Wolff (C), and three overrelaxation (O)
updates are performed. The individual updates are mixed automatically in the
program to generate the update sequence M\ M\ O\ M\ O\ M\ M\ O\ M\ C\ for
the magnetic degrees of freedom.

The random number generator is
the shift register generator R1279 defined by the recursion relation
$X_n = X_{n-p} \oplus X_{n-q}$ for $(p,q) = (1279,1063)$. Generators like
these are known to cause systematic errors in combination with the Wolff
algorithm \cite{cluerr}. However, for lags $(p,q)$ used here these errors
are far smaller than typical statistical errors. They are further reduced
by the hybrid nature of the algorithm \cite{Hybrid}.

The hybrid algorithm is well suited to explore second-order phase
transitions. However, it is unable to overcome the exponential
slowing down of the algorithms included in our hybrid scheme in the
vicinity of a first-order transition, e.g., the first-order magnetic
(i.e., superfluid) transition for higher concentrations of $^3$He
particles, i.e., for occupation  numbers $t_i=0$ in our model.
 In order to resolve this problem while keeping
the benefits of the  hybrid scheme we have embedded the hybrid Monte Carlo
method in a simulated tempering environment \cite{SimTemp}. According to
the simulated tempering idea the temperature is treated as a random
variable which performs a random walk inside a predefined temperature
interval. In our simulation this temperature interval is represented by
a discrete set of temperatures, which are spaced closely enough to allow
sufficient overlap of the corresponding energy distribution functions.
The required reweighting factors \cite{SimTemp} are estimated from short
runs, one for each pair of neighboring temperatures, and checked a posteriori
by monitoring the probability distribution of the temperatures - which
should be essentially flat -- during the production run. Deviations of up to
20\% from a flat temperature distribution are tolerated.

The Monte  Carlo scheme described above is employed for lattice sizes
$L$ between $L = 12$ and $L = 60$. For each choice of parameters we 
have performed
at least 12 blocks of $10^3$ hybrid steps for equilibration followed by
another $10^3$ hybrid steps to estimate the reweighting factors for each
pair of neighboring temperatures and finally followed by $4 \times 10^4$
hybrid steps for measurements. The measurement block is controlled by an
outer loop in which a new temperature is proposed according to the
predetermined weight factors \cite{SimTemp} after each hybrid Monte Carlo
step. Apart from standard thermodynamic quantities the distribution functions
of the total energy, the density, and the modulus of the magnetization are
monitored using histogram reweighting and extrapolation techniques
\cite{Histogram} within the measurement block. Their statistical errors are
estimated following standard procedures resulting from the statistical
independence of different measurement blocks. Unless otherwise stated
all error bars quoted in the following correspond to one standard deviation.
They are displayed only when they exceed the symbol sizes. The simulations
have been performed on DEC Alpha Workstations and Pentium III PCs.

\section{Monte Carlo results}
\label{sec:MCres}
Our primary interest in this study is to use Eq. (\ref{eq:ham})
to model $^3$He-$^4$He mixtures in the tricritical region and we therefore
restrict the numerical investigation of the statistical model described
by Eq. (\ref{eq:ham}) to the case $J = K$ for which the phase diagram
corresponding to this model Hamiltonian has the same
topology as for the liquid phases of $^3$He-$^4$He mixtures. The tricritical
point marks the onset of demixing into a spin ($^4$He) rich fluid and a
vacancy ($^3$He) rich fluid, where the spin rich fluid simultanously
exhibits long-ranged magnetic order of the XY type (superfluidity).

The phase diagram is most conveniently investigated by the inspection
of distribution functions (histograms) for various thermodynamic
quantities \cite{Nigel}.
However, the computational expense of the method described in Ref. \cite{Nigel}
in $d = 2$ is prohibitive in $d = 3$ for any appreciable system sizes. We
therefore resort to a simpler though less accurate approach which allows us
to treat larger systems and is accurate enough for our purposes. In the
following all temperatures are measured in units of the critical temperature
$T_s(0)$ of the XY model on a simple cubic lattice in $d = 3$, which is given
by $K_c \equiv J / (k_B T_s(0)) = 0.45415(5)$ \cite{TcXY}. The chemical
potential $\Delta$ is measured in units of the magnetic coupling contsant
$J$ (see Eq. (\ref{eq:ham})).

\subsection{Order parameter distribution at tricriticality}
The key feature of the $^3$He-$^4$He phase diagram is the presence of a
tricritical point. The task to locate the tricritical point for the model
Hamiltonian given by Eq. (\ref{eq:ham}) is aided by the observation that
$d = 3$ is the upper critical dimension for tricriticality. It is therefore
reasonable to assume that the distribution function of the magnetic order
parameter essentially takes the mean-field (Landau) form. We will give some a
posteriori evidence below that this assumption is indeed correct, but an
accurate numerical proof of it is beyond the scope of this paper.

The magnetic order parameter, i.e., the magnetization is defined by
\begin{equation} \label{eq:magnOP}
{\bf M} = (M_x, M_y) \equiv L^{-3} \sum_i t_i {\bf S}_i ,
\end{equation}
where $t_i = 0, 1$ characterizes the presence of $^3$He or $^4$He at lattice
site $i$ and ${\bf S}_i = (\cos \theta_i, \sin \theta_i)$ is the standard spin
variable of the XY model. In terms of the modulus $m \equiv |{\bf M}|$ of
the order parameter the distribution function $P(m)$ is assumed to take the
form
\begin{equation} \label{eq:Pofm}
P(m) = P_0 m \exp(- A m^2 - B m^4 - C m^6)
\end{equation}
according to Landau theory in the tricritical region, where the absence
of symmetry breaking fields is assumed. The parameters $A$, $B$, and $C$
essentially play the role of the Landau-Ginzburg model parameters (see
Sec. \ref{sec:LG} below) and they depend on the temperature $T$ and the
chemical potential $\Delta$ (see Eq. (\ref{eq:ham})), where $C$ is manifestly
positive, but $A$ and $B$ may change sign. For system sizes $L = 12$, 18,
24, 36, 48 and 60 simulations have been performed along various paths in the
$(T,\Delta)$ plane of the phase diagram and the data recorded for $P(m)$ have
been fitted according to Eq. (\ref{eq:Pofm}) using $P_0$, $A$, $B$, and $C$
as fit parameters. For each system size $L$ a pseudo tricritical point
$(T_t(L), \Delta_t(L))$ has been identified by the requirement $A = B = 0$
within the corresponding statistical error. It turns out, that Eq.
(\ref{eq:Pofm}) indeed captures the shape of $P(m)$ rather accurately over
several orders of magnitude for $P$ in the pseudo tricritical regime (see
below). In particular, higher powers of $m$ compatible with the symmetry such
as $m^8$ can be safely ignored. Possible logarithmic corrections to $P(m)$
\cite{LawSar} could not be identified from the numerical data unambiguously.
We will comment on other logarithmic corrections later.

\subsection{The tricritical point}
\begin{figure}[h] 
\centerline{\mbox{\epsffile{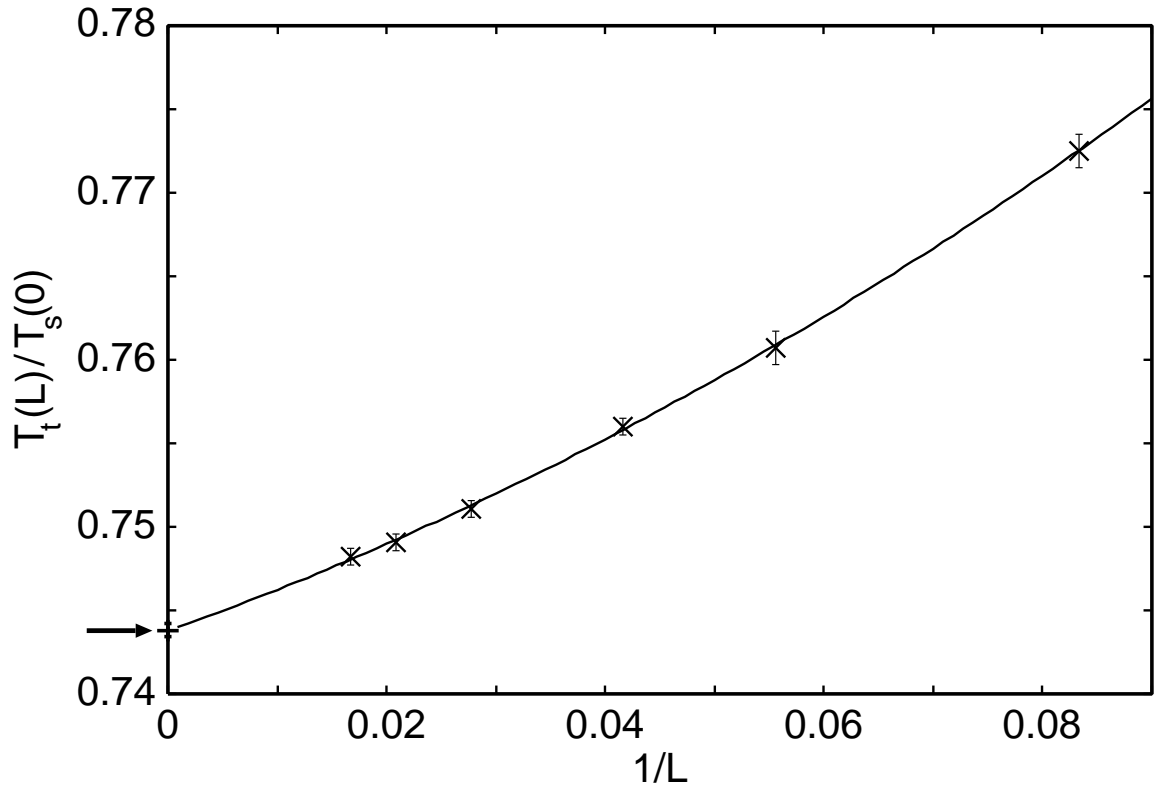}}} 
\vskip15pt 
\caption{
Pseudo tricritical temperature $T_t(L)$ $(\times)$ vs. $1/L$ measured in units
of the critical temperature $T_s(0)$ of the XY model on a s.c. lattice in
$d = 3$ for $J = K$ (see Eq. (\protect\ref{eq:ham})). Error bars correspond
to one standard deviation. The solid line shows the fit of Eq.
(\protect\ref{eq:triL}) to the numerical data. The arrow indicates the
extrapolated value $T_t$ (see main text). The reduced $\chi^2$ of the
fit is 0.15.}
\label{fig:TtL}
\end{figure}

\begin{figure}[h] 
\centerline{\mbox{\epsffile{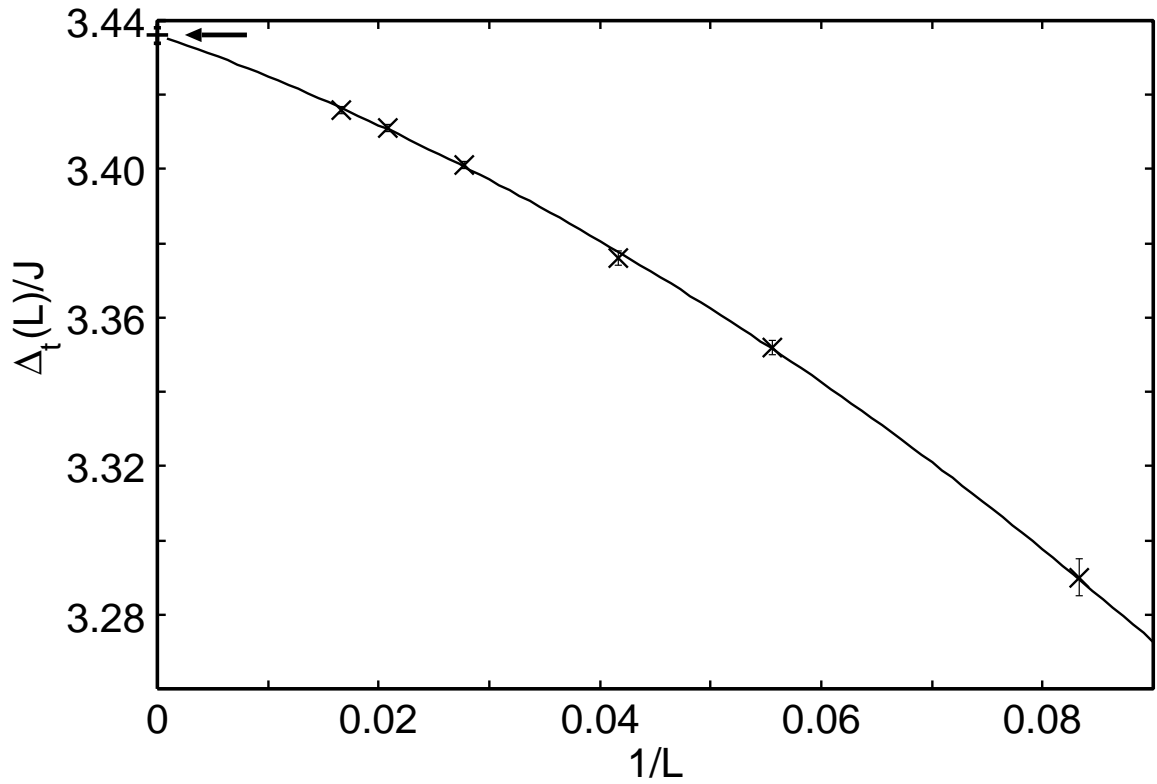}}} 
\vskip15pt 
\caption{
Pseudo tricritical chemical potential $\Delta_t(L)$ $(\times)$ vs. $1/L$
measured in units of the coupling constant $J$ (see Eq. (\protect\ref{eq:ham}))
for $J = K$. Error bars correspond to one standard deviation.
The solid line shows the fit of Eq. (\ref{eq:triL}) to the numerical
data. The arrow indicates the extrapolated value $\Delta_t$ (see main text).
The reduced $\chi^2$ of the fit is 0.42.}
\label{fig:DtL} 
\end{figure}

From the procedure outlined above we obtain a sequence of pseudo tricritical
points $(T_t(L), \Delta_t(L))$ which can be extrapolated to the bulk limit
$L \to \infty$. In order to do this one has to identify the functional form
of the $L$ dependence of the pseudo tricritical point. Within our mean-field
picture of the tricritical behavior of our model the coefficients $A$ and $B$
in Eq. (\ref{eq:Pofm}) are given by the linear combination
\begin{equation} \label{eq:BAlin}
\left(\begin{array}{c} B \\ A \end{array}\right) = {\cal M}
\left(\begin{array}{c} T - T_t \\ \Delta - \Delta_t \end{array}\right)
\end{equation}
in the vicinity of the tricritical point, where $\cal M$ is the coefficient
matrix. From Eq. (\ref{eq:BAlin}) and finite-size scaling arguments
one concludes that for sufficiently large $L$, $T_t(L) - T_t$ and
$\Delta_t(L) - \Delta_t$ are governed by a linear combination of $L^{-d_A}$
and $L^{-d_B}$, where $d_A$ and $d_B$ are the scaling dimensions of the
parameters $A$ and $B$ in Eq. (\ref{eq:Pofm}) given by $d_A = 2$ and $d_B = 1$
apart from logarithmic corrections \cite{LawSar}. We therefore arrive at
the following functional form of $T_t(L)$ and $\Delta_t(L)$:
\begin{eqnarray} \label{eq:triL}
T_t(L) &=& T_t + \frac{t_1}{L} + \frac{t_2}{L^2} \nonumber \\
\Delta_t(L) &=& \Delta_t + \frac{\delta_1}{L} + \frac{\delta_2}{L^2},
\end{eqnarray}
where the coefficients $t_1$, $t_2$, $\delta_1$, and $\delta_2$ can be
obtained from the inverse matrix ${\cal M}^{-1}$ and the finite-size
relations
\begin{equation}\label{eq:ABL}
A = a L^{-2} \ \mbox{and} \ B = b L^{-1}
\end{equation}
for $A$ and $B$ evaluated at $(T, \Delta) = (T_t(L), \Delta_t(L))$ for any
system size $L$. The coefficients $a$ and $b$ are nonuniversal metric factors.
Eq.(\ref{eq:triL}) is used to fit the numerical data for $T_t(L)$ and
$\Delta_t(L)$ in order to obtain an estimate for the location of the
tricritical point. Logarithmic corrections as given in Ref. \cite{LawSar}
can be included in Eq. (\ref{eq:triL}), but they are omitted here because the
quality of fit does not change substantially when they are included. The
results are shown in Figs. \ref{fig:TtL} and \ref{fig:DtL}. The finite-size
behavior of $T_t(L)$ and $\Delta_t(L)$ is accurately captured by Eq.
(\ref{eq:triL}). Both the coefficients $t_1$, $t_2$ and $\delta_1$, $\delta_2$
have the same sign and the second coefficient is substantially larger
that the first one in both cases. Therefore both coefficients must be kept
in order to obtain an acceptable fit. The quality of fit can be measured
in terms of the reduced $\chi^2$ which is 0.15 in Fig. \ref{fig:TtL} and
0.42 in Fig. \ref{fig:DtL}. We thus obtain the extrapolated values
\begin{equation} \label{eq:tripoint}
T_t / T_s(0) = 0.7438(4), \quad \Delta_t / J = 3.436(2)
\end{equation}
as our estimate for the location of the tricritical point, where the
statistical uncertainty affects the last digit by the amount given in
parenthesis. In these units the coefficients in Eq. (\ref{eq:triL}) are
given by
\begin{eqnarray} \label{eq:TDcoeff}
t_1 / T_s(0) = 0.23 \pm 0.02 &\ ,\ & t_2 / T_s(0) = 1.34 \pm 0.24\ , \\
\delta_1 /J = -1.05 \pm 0.11 &\ ,\ & \delta_2 / J = -8.5 \pm 1.4\ . \nonumber 
\end{eqnarray}
Another aspect of Eq. (\ref{eq:triL}) is field mixing \cite{Nigel}, because
the finite-size corrections $L^{-1}$ and $L^{-2}$ are uniquely related to
the coefficients (scaling fields) $B$ and $A$ in Eq. (\ref{eq:Pofm}),
respectively. In the vicinity of the tricritical point one therefore obtains
from Eqs. (\ref{eq:BAlin}), (\ref{eq:triL}), and (\ref{eq:ABL}) by a matrix
inversion
\begin{equation} \label{eq:fieldmix}
\left(\begin{array}{c} B/b \\ A/a \end{array}\right)
= \frac{1}{t_1 \delta_2 - t_2 \delta_1}
\left(\begin{array}{rr} \delta_2 & -t_2 \\ -\delta_1 & t_1 \end{array} \right)
\left(\begin{array}{c} T - T_t \\ \Delta - \Delta_t \end{array}\right).
\end{equation}
According to our mean field picture of tricriticality in $d=3$ the
coexistence line $\Delta = \overline{\Delta}(T)$ in the vicinity of $T=T_t$
should be associated with the line $B = \mbox{const.} = 0$ in the vicinity
of $A = 0$. If we linearize the coexistence line near the tricritical point
according to
\begin{equation} \label{eq:coex}
\overline{\Delta}(T) = \Delta_t + \Delta'_t \left(T - T_t\right)
\end{equation}
we obtain from Eqs. (\ref{eq:TDcoeff}) and (\ref{eq:fieldmix}) for the slope
$\Delta'_t$ at the tricritical point
\begin{equation} \label{eq:Dprime}
\Delta'_t = \delta_2 / t_2 = (6.4 \pm 2.2) J / T_s(0) .
\end{equation}
Despite its large statistical error this result serves as a valuable guideline
for the further exploration of the phase diagram.

\begin{figure}[h] 
\centerline{\mbox{\epsffile{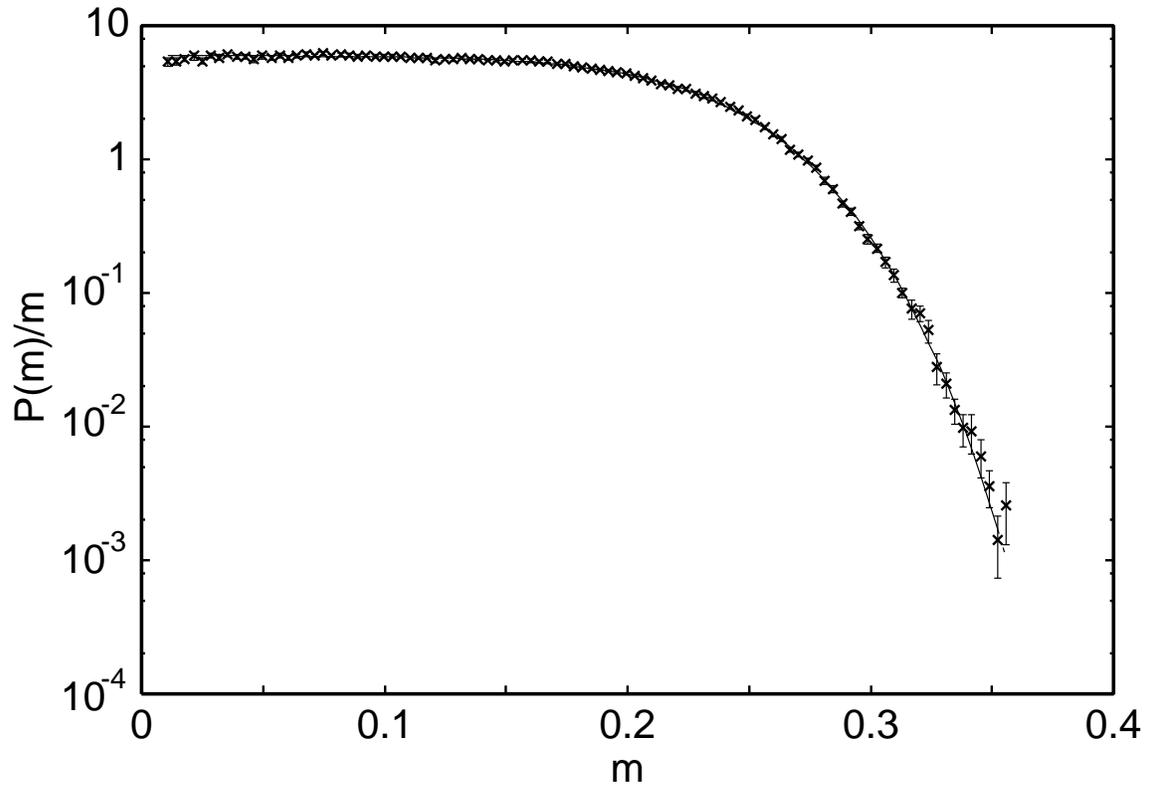}}} 
\vskip15pt
\caption{Least square fit of Eq. (\protect\ref{eq:Pofm}) (solid line) to
the simulation data for $P(m)$ for $L = 36$ at $T = T_t(36)$ and $\Delta =
\Delta_t(36)$ ($\times$) corresponding to $A = B = 0$. All data points
except very few are connected by the fit function within their error bars.
The reduced $\chi^2$ of the fit is 0.71.}
\label{fig:Pm36}
\end{figure}

The foundation of the above estimates is the quality of the fits of Eq.
(\ref{eq:Pofm})
to the measured order parameter distribution functions. We illustrate the
quality of these fits in Fig. \ref{fig:Pm36} for
$L = 36$ at the corresponding pseudo tricritical point $T = T_t(36)$ and
$\Delta = \Delta_t(36)$. The shape of the distribution function $P$ is
essentially captured by Eq. (\ref{eq:Pofm}) over more than three orders of
magnitude. The parameters $A$ and $B$ vanish within their statistical
errors. The reduced $\chi^2$ of the fit is 0.71. If $A = B = 0$ is enforced,
i.e, the fit is performed only with the parameters $P_0$ and $C$, the reduced
$\chi^2$ increases to 0.92. For all other system sizes investigated the
situation is similar. We will return to the finite size behavior of $P(m)$
after the discussion of finite-size scaling.

\subsection{Finite-size scaling}
\begin{figure}[h] 
\centerline{\mbox{\epsffile{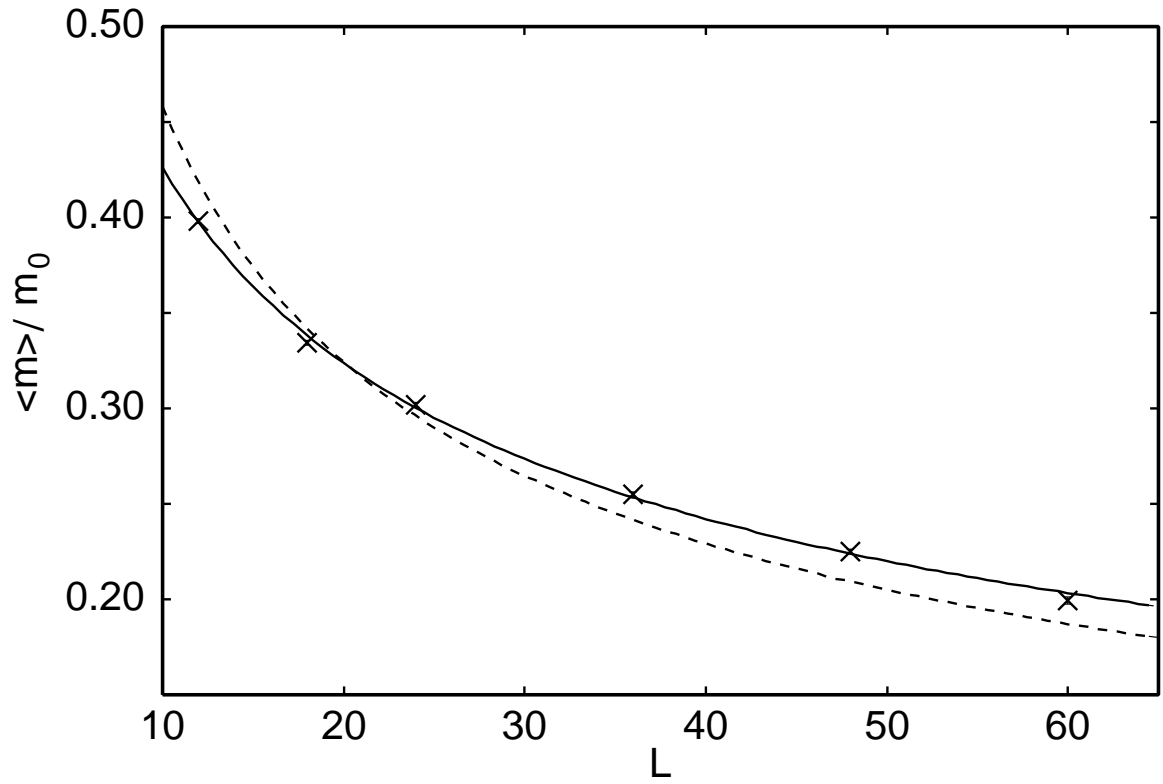}}} 
\vskip15pt
\caption{Least square fit of Eq. (\protect\ref{eq:mCXL}) (solid line) to
the simulation data for $\langle m \rangle$ ($\times$). A fit to pure
mean field behavior is shown for comparison (dashed line). Data and fit
are normalized to the amplitude $m_0$ and $l_0 = 1.3 \pm 0.3$. For $L=18$
and $L=60$ the data points deviate from the fit curve (solid line) by an
amount larger than the statistical error.}
\label{fig:mL}
\end{figure}

\begin{figure}[h] 
\centerline{\mbox{\epsffile{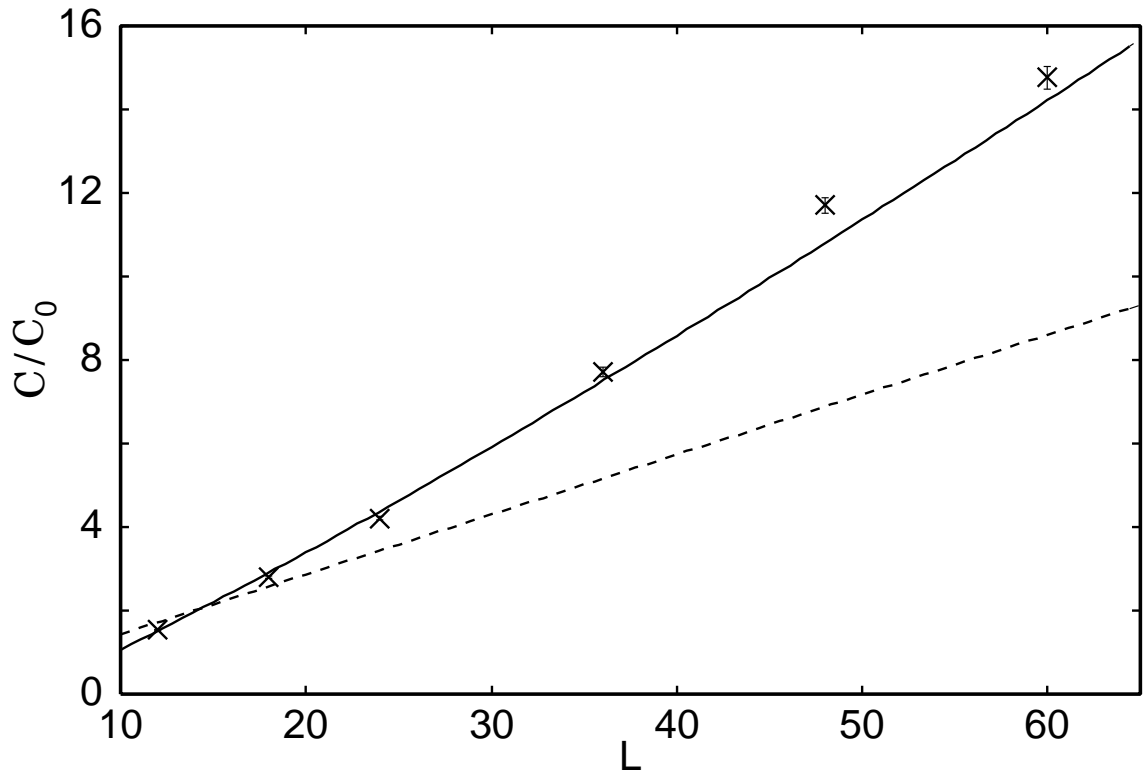}}} 
\vskip15pt
\caption{Least square fit of Eq. (\protect\ref{eq:mCXL}) (solid line) to
the simulation data for the specific heat $\cal C$ ($\times$). A fit to pure
mean field behavior is shown for comparison (dashed line). Inclusion of a
background contribution to $\cal C$ as an additional fit parameter does not
improve the fit. Data and fit are normalized to the amplitude ${\cal C}_0$
and $l_0 = 6.3 \pm 0.5$. The deviations from the expected behavior (solid
line) for larger systems may be due to the vicinity of the first-order
demixing transition.}
\label{fig:CL}
\end{figure}

\begin{figure}[h] 
\centerline{\mbox{\epsffile{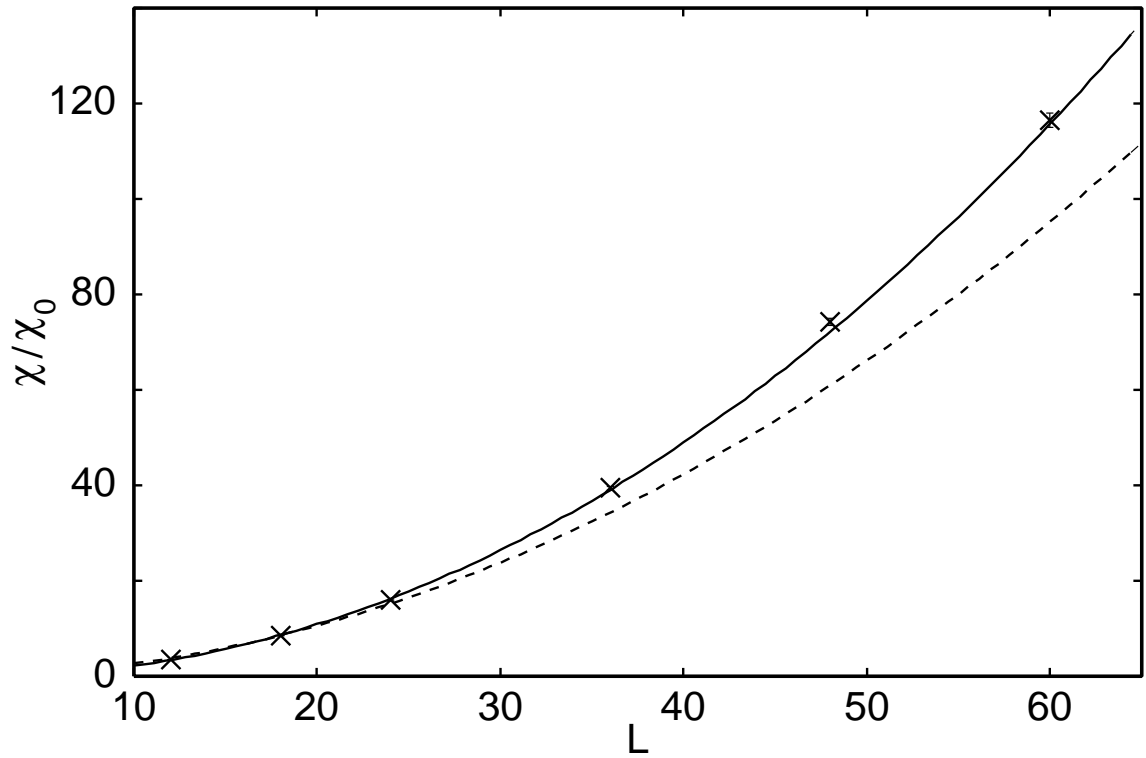}}} 
\vskip15pt
\caption{Least square fit of Eq. (\protect\ref{eq:mCXL}) (solid line) to
the simulation data for magnetic susceptibility $\cal X$ ($\times$). A fit
to pure mean field behavior is shown for comparison (dashed line). Data and
fit are normalized to the amplitude ${\cal X}_0$ and $l_0 = 6.2 \pm 0.4$.}
\label{fig:XL}
\end{figure}

A naive finite-size scaling ansatz for a thermodynamic quantity $X(A,B,L)$
near a tricritical point in $d = 3$ is given by (compare Eq. (\ref{eq:Pofm}))
\begin{equation} \label{eq:XABL}
X(A,B,L) = L^{-d_X} f_X(A L^2, B L) ,
\end{equation}
where $f_X(x,y)$ is the finite-size scaling function associated with the
quantity $X$ and $d_X$ is its scaling dimension. Logarithmic corrections
have been
disregarded for simplicity. For the sequence of the pseudo tricritical points
$(T,\Delta) = (T_t(L), \Delta_t(L))$ one has $A = a L^{-2}$ and
$B = b L^{-1}$ (see Eq. (\ref{eq:ABL})). In this case
one therefore expects $X$ to display the scaling behavior
\begin{equation} \label{eq:XtL}
X(a L^{-2},b L^{-1}, L) = L^{-d_X} f_X(a,b) \equiv X_0 L^{-d_X},
\end{equation}
which can be conveniently checked numerically. However, near tricritical
points in $d = 3$ one has to consider logarithmic corrections to the naive
scaling and these have been examined in Ref. \cite{LawSar}. We therefore only
quote the results corresponding to Eq. (\ref{eq:XtL}) for the average
magnetization $\langle m \rangle$, the specific heat $\cal C$ and the magnetic
susceptibility $\cal X$. One obtains
\begin{eqnarray} \label{eq:mCXL}
\langle m \rangle &=& m_0 \left(\frac{L}{l_0}\right)^{-1/2}
\left(\ln \frac{L}{l_0} \right)^{1/4} , \nonumber \\
{\cal C} &=& {\cal C}_0 \frac{L}{l_0}
\left(\ln \frac{L}{l_0} \right)^{1/2} , \\
{\cal X} &=& {\cal X}_0 \left(\frac{L}{l_0}\right)^2
\left(\ln \frac{L}{l_0} \right)^{1/4} , \nonumber
\end{eqnarray}
where the nonuniversal amplitudes $m_0$, ${\cal C}_0$, ${\cal X}_0$ and the
length scale $l_0$ are used as fit parameters. The corresponding results
are summarized
in Figs. \ref{fig:mL} - \ref{fig:XL}. The data are compatible with the
finite-size scaling behavior given by Eq. (\ref{eq:mCXL}) (solid lines).
The logarithmic corrections turn out to be essential. Disregarding
these leads to pure mean field behavior which is not compatible with the
data (dashed lines). For the specific heat displayed in Fig. \ref{fig:CL}
deviations from the expected behavior occur for larger system sizes $L = 48$
and $L = 60$. These may be due to the proximity to the first-order demixing
transition, which is characterized by a finite latent heat. Including a
finite background contribution to the specific heat as an additional fit
parameter does not improve the fit. In particular, the attempt to fit pure
mean field behavior to the data shown in Fig. \ref{fig:CL} leads to a negative
value for the background specific heat which is inconsistent with
thermodynamics. The susceptibility shown in Fig. \ref{fig:XL} appears to agree
with the expectation for all system sizes whereas the average magnetization
shown in Fig. \ref{fig:mL} displays a deviation for $L = 60$. The
values of $l_0$ obtained from the fits shown in Figs. \ref{fig:CL}
and \ref{fig:XL} are consistent ($l_0 = 6.3 \pm 0.5$ and $l_0 = 6.2 \pm 0.4$,
respectively) whereas $l_0 = 1.3 \pm 0.3$ obtained from $\langle m \rangle$
according to Fig. \ref{fig:mL} deviates strongly from the aforementioned
estimates for $l_0$. One of the reasons may be that $\langle m \rangle$
depends rather weakly on $L$ and $l_0$ as compared to $\cal C$ and
$\cal X$. Therefore the estimation of $l_0$ from $\langle m \rangle$ is more
susceptible to statistical or systematic errors in the magnetization data.
The large relative error of the actual estimate $l_0 = 1.3 \pm 0.3$ seems to
indicate this. Corrections to the leading asymptotic behavior given by Eq.
(\ref{eq:mCXL}), which cannot be taken into account on our current data basis,
may therefore also play a role.

\begin{figure}[h] 
\centerline{\mbox{\epsffile{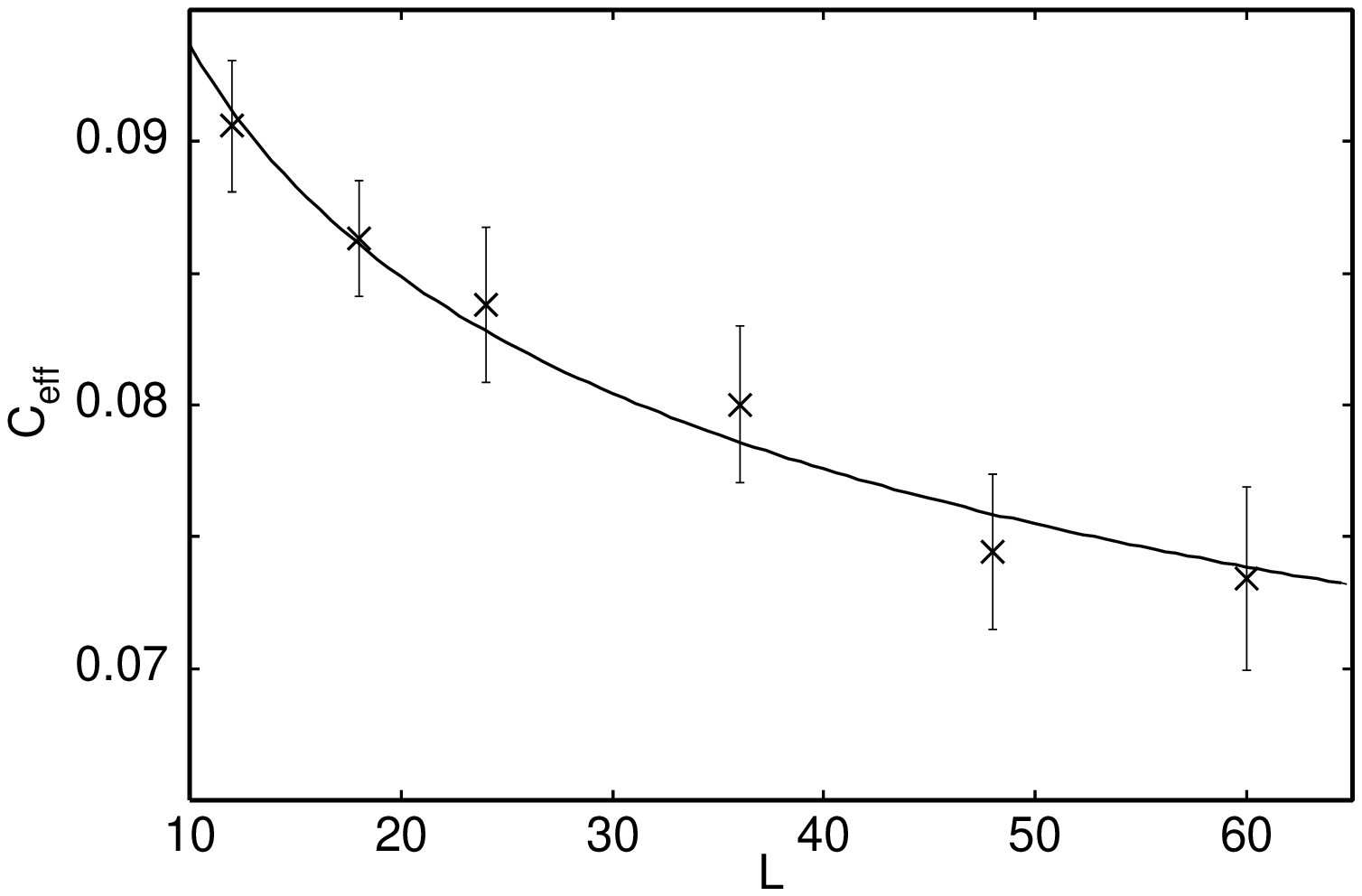}}} 
\vskip15pt
\caption{Effective coupling parameter $C_{eff}$ according to Eq.
(\protect\ref{eq:Ceff}) ($\times$) and a least square fit of Eq.
(\protect\ref{eq:Cflow}) to the data (solid line). The observed decrease
of $C_{eff}$ is compatible with the logarithmic behavior of a dangerous
irrelevant variable at the upper critical dimension. The reduced $\chi^2$
of the fit is 0.16.
}
\label{fig:CeffL}
\end{figure}

The scaling behavior of the order parameter distribution function $P(m)$
within the scope of Eq. (\ref{eq:Pofm}) is determined by the finite-size
behavior of the parameter $C$. In order to compensate the finite-size
effects induced by $\langle m \rangle$ we define the effective coupling
parameter
\begin{equation} \label{eq:Ceff}
C_{eff} \equiv C \langle m \rangle^6,
\end{equation}
where $C$ is taken from the fit of Eq. (\ref{eq:Pofm}) to the distribution
function data along the sequence of the pseudo tricritical points
($A = B = 0$) and $\langle m \rangle$ is taken from the fit of Eq.
(\ref{eq:mCXL}) to the magnetization data. The numerical result for $C_{eff}$
according to Eq. (\ref{eq:Ceff}) is displayed in Fig. \ref{fig:CeffL} which
shows a slow but systematic decrease of $C_{eff}$ with the system size.
According to the renormalization group theory of tricritical behavior
$C_{eff}$ should play the role of the coupling constant at tricriticality,
which is a dangerous irrelevant variable in $d = 3$ \cite{LawSar}. We
therefore expect the finite-size behavior \cite{LawSar}
\begin{equation} \label{eq:Cflow}
C_{eff}(L) = \left(c_0 + c_1 \ln \frac{L}{l_0} \right)^{-1}
\end{equation}
for the effective coupling parameter, where $l_0 = 6.2$ is taken from Fig.
\ref{fig:XL} and $c_0$ and $c_1$ serve as fit parameters to the data. The
solid line in Fig. \ref{fig:CeffL} displays this fit with $c_0 = 9.9 \pm 0.2$
and $c_1 = 1.6 \pm 0.1$ and demonstrates, that
the expected behavior according to Eq. (\ref{eq:Cflow}) is consistent with
the data.

The degree of agreement between the finite-size scaling behavior observed
and expected may also be considered as an a posteriori confirmation that the
sequence of the pseudo tricritical points gives a reasonable estimate for the
location of the tricritical point. However, some confirmation from a
different source would still be desirable.

\subsection{Other distribution functions and the cumulant method}
In order to locate the first-order coexistence line one may inspect the
distribution function $P(n)$ of the particle density
\begin{equation} \label{eq:dens}
n \equiv L^{-3} \sum_i t_i .
\end{equation}
Near a first-order demixing transition $P(n)$ displays two peaks,
one at a higher density corresponding to the spin ($^4$He) rich liquid
and one at a lower density corresponding to the vacancy ($^3$He) rich
liquid. At the tricritical point the two peaks coalesce and they separate
increasingly as one follows the two-phase coexistence line towards lower
temperatures. As a criterion to locate the coexistence line one may
demand that the ratio of the statistical weights of the two liquids, i.e.,
the ratio of the areas under the respective peaks of $P(n)$, should not
depend on temperature. However, this criterion is only approximate, because
a priori it is not clear what the value of the weight ratio should be.
An accurate criterion can be obtained from the evaluation of the field mixing
\cite{Nigel} (see Eq. (\ref{eq:fieldmix})). From thermodynamic considerations
one may determine a linear combination of particle and energy density
such that the corresponding distribution function is symmetric on
the coexistence line \cite{Nigel}. The generally unknown value of the
weight ratio then has to be unity at coexistence. In principle, one may
determine the correct mixing ratio of the densities from Eqs.
(\ref{eq:TDcoeff}) and (\ref{eq:fieldmix}). However, the statistical
uncertainties of the coefficients given by Eq. (\ref{eq:TDcoeff}) are too
large for this purpose and their accurate evaluation is beyond the scope
of this work.

\begin{figure}[h] 
\centerline{\mbox{\epsffile{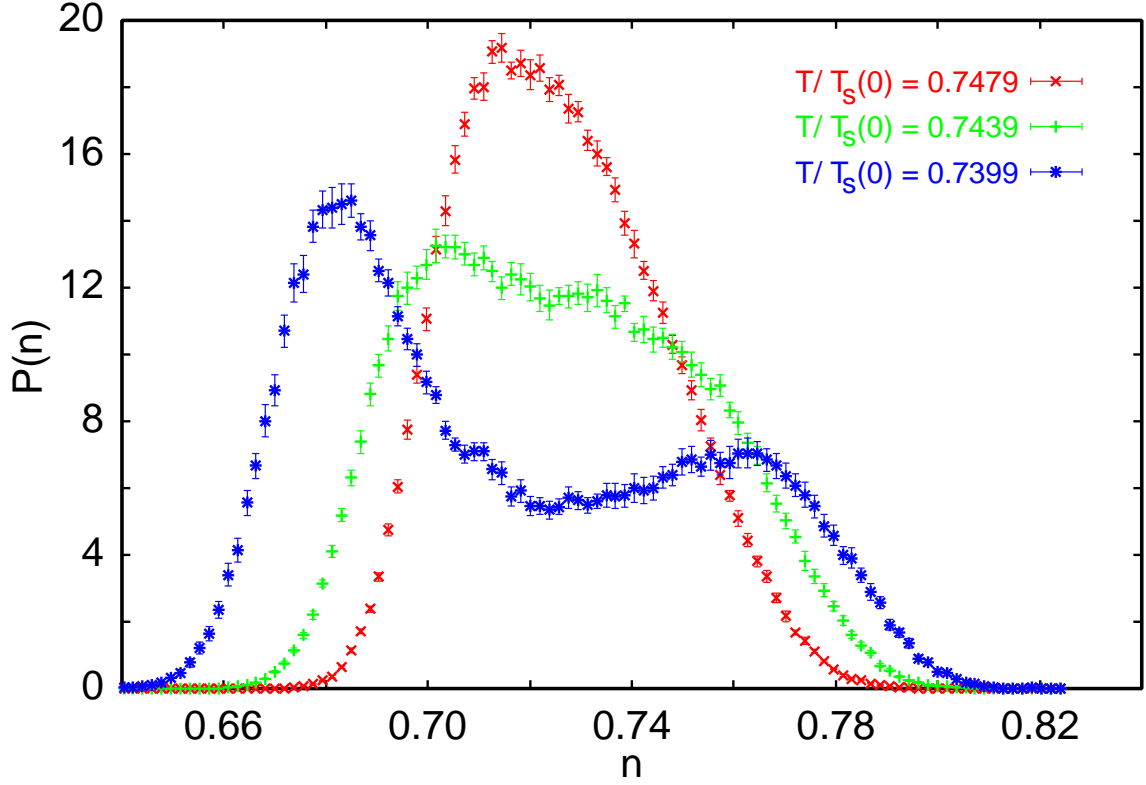}}} 
\vskip15pt
\caption{Particle density distribution $P(n)$ for three temperatures along
a straight path given by Eq. (\protect\ref{eq:coex}) in the tricritical
region as proposed by Eq. (\protect\ref{eq:tripoint}). The temperatures
chosen are $T/T_s(0) = 0.7479$ ($\times$), $T/T_s(0) = 0.7439$ (+), and
$T/T_s(0) = 0.7399$ ($*$). The parameters of Eq. (\protect\ref{eq:coex})
are $T_t/T_s(0) = 0.7439$, $\Delta_t/J = 3.438$, and $\Delta'_t = 5.0
J/T_s(0)$.}
\label{fig:PofnT}
\end{figure}

For the vector BEG model the task of locating the first-order
coexistence line is aided by the observation that the spin rich fluid
displays long-ranged XY type (superfluid) order when the demixing
transition occurs. Apart from the particle density distribution (see
Eq. (\ref{eq:dens})) we therefore also observe the distribution function
of the magnetic energy density $\varepsilon_m$ defined by
\begin{equation} \label{eq:Emagn}
\varepsilon_m \equiv -J L^{-3}\sum _{<ij>}t_i t_j\cos (\theta _i-\theta _j).
\end{equation}
For our choice $J = K$ the demixing transition will also be indicated by
a double peak structure of the distribution function $P(\varepsilon_m)$.
Note that this will no longer be the case for sufficiently large $K > J$,
for which the demixing transition precedes the onset of long-ranged magnetic
order. By monitoring both distribution functions along
various paths in the $(T,\Delta)$ plane of the phase diagram and by applying
the constant weight ratio criterion to both we have redetermined the
slope of the coexistence line in the vicinity of $(T_t,\Delta_t)$ (see
Eq. (\ref{eq:tripoint})) and found
\begin{equation} \label{eq:Dprime1}
\Delta'_t = (5.0 \pm 0.1) J / T_s(0) .
\end{equation}
Note that the new estimate given by Eq. (\ref{eq:Dprime1}) is consistent
with the previous one given by Eq. (\ref{eq:Dprime}). We furthermore observe,
that the two peaks indeed merge into a single but broader one very close to
the estimate of $(T_t,\Delta_t)$ given by Eq. (\ref{eq:tripoint}). We
illustrate this
for $P(n)$ in Fig. \ref{fig:PofnT} for $L = 36$ along a straight path in
the phase diagram according to Eq. (\ref{eq:coex}) for the choice $T_t/Ts(0)
= 0.7439$, $\Delta_t/J = 3.438$ and $\Delta'_t = 5.0 J /T_s(0)$ (see Eq.
(\ref{eq:Dprime1})) for three temperatures. In order to obtain a clear
double peak structure in $P(n)$ including the transition from and to a
single peak along the chosen path a substantial amount of fine tuning for
both $T_t$ and $\Delta_t$ is required even for moderate system sizes. It is
therefore very comforting that the values for $T_t$ and $\Delta_t$ required
to obtain Fig. \ref{fig:PofnT} are already within the error bars of the
extrapolation estimate of the tricritical point given by Eq.
(\ref{eq:tripoint}). The structure of $P(\varepsilon_m)$ essentially looks
the same so we do not reproduce it here.

The location of the tricritical point along the coexistence line can be
identified by a cumulant crossing of a suitably chosen density \cite{Nigel}.
As we have not evaluated the field mixing here we use the cumulants of the
magnetic energy density defined by Eq. (\ref{eq:Emagn}) in order to
investigate the cumulant crossing along the straight path used in Fig.
\ref{fig:PofnT}. We define the Binder cumulant ratio for $\varepsilon_m$ by
\begin{equation} \label{eq:cumEmagn}
U_{\varepsilon_m} \equiv 1 -
\frac{\langle (\varepsilon_m - \langle\varepsilon_m \rangle)^4\rangle}
{3\langle (\varepsilon_m - \langle\varepsilon_m \rangle)^2\rangle^2} .
\end{equation}
The cumulant $U_{\varepsilon_m}$ as function of temperature for different
system sizes is shown in Fig. \ref{fig:cumuE}. A unique crossing cannot be
identified. However, the various crossings occur roughly where they are
expected according to Eq. (\ref{eq:tripoint}). If the smallest system
$L = 12$ is excluded the crossings are located in the temperature interval
$0.743 < T/T_s(0) < 0.747$ which includes the estimate given by Eq.
(\ref{eq:tripoint}) near the lower bound. The crossings for larger
systems tend to occur at lower temperatures. One of the reasons that a
unique crossing does not occur is that both $\varepsilon_m$ and $n$
contain corrections to the order parameter of the demixing transition which
can only be eliminated by a properly chosen linear combination of these
quantities \cite{Nigel}. A second reason is insufficient fine tuning for
larger systems, which becomes visible in the nonmonotonic behavior of
$U_{\varepsilon_m}$ for $L = 36$ at lower temperatures which leads to
a second intersection with $U_{\varepsilon_m}$ for $L = 24$. Despite
the quantitative shortcomings of Fig. \ref{fig:cumuE} the investigation of
the cumulant crossing in combination with the other evidence presented above
provides some independent confirmation that our initial assertion about the
shape of the tricritical order parameter distribution function according
to Eq. (\ref{eq:Pofm}) is correct within the accuracy needed for the purpose
of this work.

Considerations in the spirit of Landau mean-field theory have played a major
role for the analysis of our numerical data in the tricritical region. We
therefore now turn to a detailed derivation of the Ginzburg - Landau
model in the tricritical region of the vector BEG model.

\section{Landau-Ginzburg model for $^3$He-$^4$He mixtures.}
\label{sec:LG}

In this section  we derive a two-parameter  Landau-Ginzburg (LG) model 
describing  bulk  $^3$He-$^4$He  mixtures near tricriticality.
This derivation follows  the construction of the $\phi ^4$ model
for the standard  critical phenomena  from the Ising model.

\subsection{Derivation of the model}
\label{subsec:der}
Our starting point is the modified VBEG  model for which instead of continuum
orientations of the  spin  vector ${\bf S}_i=(\cos\theta_i,\sin\theta_i)$  we consider  $L$ discrete orientations $\theta_i^{(l)}=2\pi l/L,~~ l=1,\cdots,L$,
uniformly distributed over the unit circle with $L\to \infty$.
With each orientation
 ${\bf S}^{(l)}_i=(\cos\theta_i ^{(l)},\sin\theta_i^{(l)})$   at
the site $i$ we associate the  density
$ t_{i,l}$
with  discrete values 0 or 1.
The total density of  $^4$He  at
the site $i$ is 
\begin{equation}
\label{eq:totden}
t_i=\sum _{l=1}^L t_{i,l}.
\end{equation}
As in the  VBEG model we consider the close-packing case 
in which a  $^3$He atom at site $i$ corresponds  to $t_i=0$ and a
 $^4$He  atom to $t_i=1$. Thus a lattice site $i$  is either occupied
by a $^3$He atom ($t_i=0$) or a $^4$He atom associated with one of the $L$ orientatons ($t_{i,l}=1$ for $l=l_0, t_{i,l}=0$ otherwise, so that $t_i=\sum _{l=1}^Lt_{i,l}=1$).
The Hamiltonian of this effectively  $(L+1)$- component mixture has the form
\begin{equation}
\label{eq:modham}
{\cal H}=-\sum _{<ij>}\{ J\sum _{l=1}^L\sum_{l'=1}^Lt_{i,l}t_{j,l'}{\bf S}_i^{(l)}\cdot{\bf S}_j^{(l')} +Kt_it_j\}+\Delta\sum _it_i,
\end{equation}
where ${\bf S}_i^{(l)}\cdot{\bf S}_j^{(l')}=\cos(\theta_i ^{(l)}-\theta_j^{(l')})$.

In the Landau-Ginzburg model the effective Hamiltonian, depending on the local
order-parameter fields, is obtained as a result of  coarse-graining
procedures. The  procedure which gives an
exact functional representation for the partition function 
for  the corresponding microscopic Hamiltonian  is the Hubbard-Stratonovitch
 transformation. The application of this method, however, is limited to
microscopic Hamiltonians that can be expressed as a quadratic form. 
Here we use another approach~\cite{ciach:96:0}.

Within the standard mean-field treatment of the lattice gas mixture defined by
the Hamiltonian in Eq.(\ref{eq:modham}) the ensemble-avaraged occupancy
 $\rho _{i,l}= <t_{i,l}>$
of the site $i$ is obtained by minimizing the grand canonical function 
\begin{equation}
\label{eq:omegaMF}
\Omega ^{MF}(\rho_{i,l})={\cal H}(\rho_{i,l})+\sum _i f_{id}(\rho_{i,l})
\end{equation}
at fixed $T$ and $\Delta$. The resulting minimum of $\Omega^{MF}$ equals the equilibrium grand potential $\Omega _0$.
The ideal or noninteracting free-energy density for 
a $(L+1)$ component mixture on the close-packed lattice is~\cite{bell:89:0}
\begin{equation}
\label{eq:fideal}
f_{id}(\rho_{i,l})= k_BT\{(1-\rho_i)\ln (1-\rho_i)+\sum_{l=1}^L\rho_{i,l}\ln \rho_{i,l}\}, 
\end{equation}
with
 $\rho _i=\sum _{l=1}^L\rho_{i,l}$.

In the following we shall treat $\rho _i$ and $\rho_{i,l}$  as coarse-grained 
order parameter fields and adopt the mean field grand canonical function
$\Omega^{MF}$ form for the free energy of a particular local configuration of
the order parameter.
In the spatially uniform and  orientationally disordered phase the 
equilibrium values  of $\rho_{i,l}$ are constant and
 denoted by $<\rho_{i,l}>=Q/L$ so that  $<\rho_i>=Q$.
The actual values fluctuate around these mean values:
\begin{equation}
\label{eq:ordpar1}
\rho_i=Q+\phi _i 
\end{equation}
\begin{equation}
\label{eq:ordpar2}
\rho_{i,l}=\frac{Q}{L}+\frac{\phi_i }{L}+\Delta \rho_{i,l}=\frac{\rho_i}{L}+\Delta \rho_{i,l},
\end{equation}
which implies  $<\phi_i>=0 $ and  $\sum _l \Delta \rho_{i,l}=0$
even without taking  the thermal average. 
The fluctuation of the density $\rho_{i,l}$ at the  site $i$  consists  of an
orientationally  uniform part  $\phi_i /L$ related to the fluctuation $\phi _i$
 of the total  $^4$He density, which is
the same for  all orientations, and a contribution  $\Delta \rho_{i,l}$
as  an excess density
of $^4$He in the  orientation  ${\bf S}_i^{(l)}$.

Assuming   $\Delta \rho_{i,l}$ and $\phi _i$ to be small
 we expand $\Omega ^{MF}(\rho_i,\rho_{i,l})$ in power series 
 of the fluctuation fields  $\Delta\rho_{i,l}$  and   $\phi _i$
 about the equilibrium value
 $\Omega _0^{MF}(\rho_i=Q,\rho_{i,l}=Q/L)$.
Since we aim for deriving an effective Hamiltonian 
describing  bulk  $^3$He-$^4$He  mixtures near tricriticality,
in the expansion  we keep
terms to  the sixth order in $\Delta \rho _{i,l}$ and to
quadratic order in $\phi_i$.
A standard Taylor expansion  gives
\begin{equation}
\label{eq:omega2}
\Omega^{MF}(\rho _i,\rho _{i,l})-\Omega_0^{MF}(Q,Q/L)=\sum _{j=2}^6\Omega_j^{MF}(\phi _i,\Delta\rho_{i,l}).
\end{equation}
The contribution linear  in the fluctuation fields $\Omega _1^{MF}$  vanishes
 since we expand $\Omega^{MF}$ around its minimum.
The other terms are:
\begin{eqnarray}
\label{eq:omega3}
\Omega_2^{MF}(\phi _i,\Delta \rho_{i,l})&=&-\sum _{<i,j>}\{K\phi_i\phi_j+J\sum _{l=1}^L\sum _{l'=1}^L\Delta \rho_{i,l}\Delta \rho_{j,l'}{\bf S}_i^{(l)}\cdot{\bf S}_j^{(l')}\} \nonumber \\
&+&\frac{k_BT}{2}\sum _i\{\frac{1}{Q(1-Q)}\phi _i^2+\sum _{l=1}^L\frac{L}{Q}(\Delta\rho_{i,l})^2\},
\end{eqnarray}  
\begin{equation}
\label{eq:omega4}
\Omega _3^{MF}(\phi _i,\Delta \rho_{i,l})= -\frac{k_BT}{2}\sum _i\sum _{l=1}^L\{ \frac{L}{Q^2}\phi _i(\Delta\rho_{i,l})^2+\frac{1}{3}\frac{L^2}{Q^2}(\Delta\rho_{i,l})^3\}, 
\end{equation}  
\begin{equation}
\label{eq:omega5}
\Omega _4^{MF}(\phi _i,\Delta \rho_{i,l})= \frac{k_BT}{2}\sum _i \sum _{l=1}^L\{\frac{2}{3}\frac{L^2}{Q^3}\phi_i(\Delta\rho_{i,l})^3+\frac{L}{Q^3}\phi_i^2(\Delta\rho_{i,l})^2+\frac{1}{6}\frac{L^3}{Q^3}(\Delta\rho_{i,l})^4\}, 
\end{equation}  
\begin{equation}
\label{eq:omega6}
\Omega _5^{MF}(\phi _i,\Delta \rho_{i,l})= -\frac{k_BT}{2}\sum _i \sum _{l=1}^L\{\frac{1}{2}\frac{L^3}{Q^4}\phi_i(\Delta\rho_{i,l})^4+\frac{L^2}{Q^4}\phi_i^2(\Delta\rho_{i,l})^3+\frac{1}{10}\frac{L^4}{Q^4}(\Delta\rho_{i,l})^5\}, 
\end{equation}  
\begin{equation}
\label{eq:omega7}
\Omega _6^{MF}(\phi _i,\Delta \rho_{i,l})= \frac{k_BT}{2}\sum _i \sum _{l=1}^L\{\frac{2}{5}\frac{L^4}{Q^5}\phi_i(\Delta\rho_{i,l})^5+\frac{L^3}{Q^5}\phi_i^2(\Delta\rho_{i,l})^4+\frac{1}{15}\frac{L^5}{Q^5}(\Delta\rho_{i,l})^6\}. 
\end{equation}

The excess density in  the orientation  ${\bf S}_i^{(l)}$
is a periodic function of $l$ with period $L$. Therefore, it
can be expanded into a discrete  Fourier series
\begin{equation}
\label{eq:four}
\Delta\rho_{i,l}=\frac{Q}{2L}\sum_{k=1}^{L-1}u_{i,k}e^{i(2\pi/L)kl},
\end{equation}
where we have chosen $(Q/2L)$ as a normalization constant.
The term corresponding to $k=0$ is excluded from the expansion
due to the constrain $\sum _{l=1}^L\Delta\rho_{i,l}=0$.
Since $\Delta\rho_{i,l}$ is a real function, the Fourier components
$u_{i,k}$ and $u_{i,L-k}$  are related by  $ u^*_{i,k}=  u_{i,L-k}$. 
Now, we neglect higher modes in the Fourier expansion (\ref{eq:four})
and  approximate the excess density in  the orientation  ${\bf S}_i^{(l)}$
by
\begin{equation}
\label{eq:four1}
\Delta\rho_{i,l}\approx \frac{Q}{2L}\left( u_{i,1}e^{i(2\pi/L)l}+ u_{i,L-1}e^{-i(2\pi/L)l}\right).
\end{equation}
Using  $ u^*_{i,k}=  u_{i,L-k}$ and expressing
 $ u_{i,1}$ in terms of its amplitude $|u_i|$ and a phase $\zeta_i$ we write
Eq.~(\ref{eq:four1}) as
\begin{equation}
\label{eq:fou2}
\Delta\rho_{i,l}\approx \frac{Q}{2L}|u_i|\left( e^{i(2\pi/L)l+\xi_i}+
 e^{-i((2\pi/L)l+\xi_i)}\right)=\frac{Q}{L}|u_i|\cos((2\pi/L)l+\xi_i)=\frac{Q}{L}{\bf S}_i^{(l)}\cdot {\bf u}_i,
\end{equation}
where  ${\bf u}_i\equiv |u_i|\exp(i\zeta_i)$.

The approximation  (\ref{eq:fou2})
is a 'coarse-graining' procedure  which reduces degrees of freedom.
The $L-1$ independent quantities describing the orientational degrees of freedom  are replaced by a   2-component  vector field ${\bf u}_i$.

We define the 
 effective Hamiltonian for the order parameter fields $\phi $
and  ${\bf u}$  as:
\begin{equation}
\label{eq:omega2p}
\Omega^{eff}(\phi_i,{\bf u}_i)\equiv \Omega^{MF}(\rho _i,\rho _{i,l})-\Omega_0^{MF}(Q,Q/L).
\end{equation}
Using Eq.~(\ref{eq:fou2}) in Eqs.~(\ref{eq:omega3}-\ref{eq:omega7})
we obtain
\begin{equation}
\label{eq:omaga0p}
\Omega^{eff} =\Omega_2^{eff}+\Omega_{int}^{eff}
\end{equation}
with
\begin{eqnarray}
\label{eq:omega3p}
\Omega_2^{eff}(\phi _i,{\bf u}_i)&=&-\sum _{<i,j>}\{K\phi_i\phi_j+\frac{Q^2}{L^2}J\sum _{l=1}^L\sum _{l'=1}^L({\bf S}_i^{(l)}\cdot {\bf u}_i)({\bf S}_j^{(l')}\cdot {\bf u}_j){\bf S}_i^{(l)}\cdot{\bf S}_j^{(l')}\} \nonumber \\
&+&\frac{k_BT}{2}\sum _i\{\frac{1}{Q(1-Q)}\phi _i^2+\frac{Q}{L}\sum _{l=1}^L({\bf S}_i^{(l)}\cdot {\bf u}_i)^2\},
\end{eqnarray}  
\begin{eqnarray}
\label{eq:omega4p}
\Omega _{int}^{eff}&=& \frac{k_BT}{2}\sum _i\{-\phi _i \frac{1}{L}\sum _{l=1}^L({\bf S}_i^{(l)}\cdot {\bf u}_i)^2+\phi_i^2\frac{1}{QL}\sum _{l=1}^L({\bf S}_i^{(l)}\cdot {\bf u}_i)^2+\frac{1}{6}\frac{Q}{L}\sum _{l=1}^L({\bf S}_i^{(l)}\cdot {\bf u}_i)^4 \nonumber \\
&-&\frac{1}{2}\phi_i\frac{1}{L}\sum _{l=1}^L({\bf S}_i^{(l)}\cdot {\bf u}_i)^4+\phi_i^2\frac{1}{QL}\sum _{l=1}^L({\bf S}_i^{(l)}\cdot {\bf u}_i)^4+\frac{1}{15}\frac{Q}{L}\sum _{l=1}^L({\bf S}_i^{(l)}\cdot {\bf u}_i)^6\}. 
\end{eqnarray}  
We note that in the above expression terms containing sums over all
orientations of odd powers of $({\bf S}_i^{(l)}\cdot {\bf u}_i)$ vanish.

As the next step we  take the limit $L\to \infty$. This ammounts to replacing
$\frac{1}{L}\sum _{l=1}^L$ by $ \frac{1}{2\pi}\int_0^{(2\pi)} d\theta$
and leads to the following   relations $({\bf S}_i^{(l)}=(\cos\theta_i ^{(l)},\sin\theta_i^{(l)}))$:
\begin{equation}
\label{eq:re1}
\frac{Q}{L}\sum _{l=1}^L({\bf S}_i^{(l)}\cdot {\bf u}_i)^2\to \frac{Q}{2\pi}\int_0^{2\pi} d\theta_i ^{(l)}({\bf S}_i^{(l)}\cdot {\bf u}_i)^2=\frac{Q}{2}| {\bf u}_i|^2,
\end{equation}
\begin{equation}
\label{eq:re2}
\frac{Q}{L}\sum _{l=1}^L({\bf S}_i^{(l)}\cdot {\bf u}_i)^4\to \frac{Q}{2\pi}\int_0^{2\pi}d\theta_i^{(l)}({\bf S}_i^{(l)}\cdot {\bf u}_i)^4=\frac{3Q}{8}| {\bf u}_i|^4, 
\end{equation}
\begin{equation}
\label{eq:re3}
\frac{Q}{L}\sum _{l=1}^L({\bf S}_i^{(l)}\cdot {\bf u}_i)^6\to \frac{Q}{2\pi}\int_0^{2\pi} d\theta_i^{(l)}({\bf S}_i^{(l)}\cdot {\bf u}_i)^6=\frac{5Q}{16}| {\bf u}_i|^6, 
\end{equation}
and 
\begin{eqnarray}
\label{eq:re4}
\frac{Q^2}{L^2}\sum _{l=1}^L\sum _{l'=1}^L({\bf S}_i^{(l)}\cdot {\bf u}_i)({\bf S}_j^{(l')}\cdot {\bf u}_j){\bf S}_i^{(l)}\cdot{\bf S}_j^{(l')}\nonumber \\
 & \to& \frac{Q^2}{4\pi ^2}\int _0^{2\pi} d\theta_i^{(l)}\int _0^{2\pi} d\theta_j^{(l')}({\bf S}_i^{(l)}\cdot{\bf u}_i)({\bf S}_j^{(l')}\cdot{\bf u}_j){\bf S}_i^{(l)}\cdot {\bf S}_j^{(l')}\nonumber \\
&=&\frac{Q^2}{4}{\bf u}_i\cdot {\bf u}_j.
\end{eqnarray}
Finally, we asume that the fluctuating  fields $\phi_i$ and ${\bf u}_i$ vary 
slowly on the length scale of the lattice constant $a$.
The continuum limit is obtained by considering  $a\to 0$, considering $i$ as
 a continuous variable ${\bf r}$,  and $\phi _i$ and ${\bf u}_i$
turning into  $\phi ({\bf r})$ and ${\bf u}({\bf r})$, respectively, while
keeping the total volume $V=a^3N$ fixed.
 In this limit, one has
\begin{equation}
\label{eq:l1}
 \sum _i\to a^{-3}\int d{\bf r}.
\end{equation}
For $f$ being the  smooth continuation to continuous  arguments of a function
defined on a  lattice we use the following  approximations ($a \to 0$):
\begin{equation}
\label{eq:l2}
\sum _{k=1}^df({\bf r}+a{\bf e}_k)\to df({\bf r})+a\sum_{k=1}^d\frac{\partial f}{{\partial r}_k}+\frac{a^2}{2}\sum_{k=1}^d\frac{\partial^2 f}{{\partial r}^2_k}+\cdots
\end{equation}
and
\begin{equation}
\label{eq:l3}
\sum _{k=1}^d\{f({\bf r}+a{\bf e}_k)+f({\bf r}-a{\bf e}_k)\} \to
2df({\bf r})+a^2\sum_{k=1}^d\frac{\partial^2 f}{{\partial r}^2_k}+\cdots,
\end{equation}
where ${\bf e}_k, k=1,\ldots, d$ are the unit lattice vectors.
Thus 
\begin{eqnarray}
\label{eq:l4}
\sum_{<i,j>}f_if_j&\to&\frac{1}{2}a^{-3}\int d^3{\bf r}\sum_{k=1}^df({\bf r})\{f({\bf r}+a{\bf e}_k)+f({\bf r}-a{\bf e}_k)\}\nonumber \\ 
&\to& \frac{1}{2}a^{-3}\int d^3r\{2df^2({\bf r})-a^2(\nabla f)^2\}.
\end{eqnarray}
As a result Eq. (\ref{eq:omega2p}) is replaced by
\begin{equation}
\label{eq:omegaeff}
\Omega ^{eff}=K[\Omega _G+\Omega _{int}]
\end{equation}
with the Gaussian contribution $\Omega_G$, in which the
 fields $\phi$ and ${\bf u}$ are uncoupled,
\begin{equation}
\label{eq:omegaeff2}
\Omega_G=\int d{\bf r}\{ \frac{1}{2}a_1\phi ^2+\frac{1}{2}(\nabla \phi)^2+\frac{1}{2}a_2|{\bf u}|^2+\frac{1}{2}c(\nabla {\bf u})^2\},
\end{equation}
and the interaction contribution, which couples $\phi $ and  ${\bf u}$, 
\begin{equation}
\label{eq:omegaeffin}
\Omega_{int}=\int d{\bf r}\{r_1\phi |{\bf u}|^2 +a_{12}\phi^2|{\bf u}|^2+b|{\bf u}|^4+r_2\phi|{\bf u}|^4+b_{12}\phi^2|{\bf u}|^4+e|{\bf u}|^6 \},
\end{equation}
where we have  chosen the length unit such that $a=1$.

The coupling constants  in the effective functional are given explicitly by:
\begin{equation}
\label{eq:p1}
a_1=\frac{k_BT}{K}\frac{1}{Q(1-Q)}-z,
\end{equation}
\begin{equation}
\label{eq:p2}
a_2=\frac{k_BT}{K}\frac{Q}{2}-\frac{zQ^2}{4}\frac{J}{K},
\end{equation}
\begin{equation}
\label{eq:p2a}
c=\frac{Q^2}{4}\frac{J}{K},
\end{equation}
\begin{equation}
\label{eq:p4}
r_1=-\frac{1}{4}\frac{k_BT}{K},~~~~~~~~~r_2=-\frac{3}{32}\frac{k_BT}{K},
\end{equation}
\begin{equation}
\label{eq:p3}
b=\frac{Q}{32}\frac{k_BT}{K},~~~~~~~~~~~e=\frac{Q}{96}\frac{k_BT}{K},
\end{equation}
\begin{equation}
\label{eq:p6}
a_{12}=\frac{1}{4Q}\frac{k_BT}{K},~~~~~~~~~~ b_{12}=\frac{3}{16Q}\frac{k_BT}{K}.
\end{equation}
Equations ~(\ref{eq:omegaeff2}), (\ref{eq:omegaeffin}) and (\ref{eq:p1}-\ref{eq:p6}) define the  Landau-Ginzburg model for $^3$He-$^4$He mixtures
in terms of thermodynamical quantities
and two parameters, J and K, characterizing the system.

\subsection{$\lambda$-line and tricritical point}
\label{subsec:p}
In this subsection we determine the  $\lambda$-line and a tricritical 
point within   mean-field theory for  the LG  model
 derived in the preceding subsection. To this end we consider 
  spatially uniform order parameter fields. 

Mean-field theory amounts to  approximating the thermodynamic 
free  energy  by  the minimum of the effective Hamiltonian
 which corresponds to the saddle point path contribution to the partition function:
\begin{equation}
\label{eq:maGL}
\beta F_{MF}=\mbox{min}_{\phi,{\bf u}}\beta\Omega ^{eff}[\phi,{\bf u}].
\end{equation}
The mean-field solution for $\phi({\bf r})$ is determined by
\begin{equation}
\label{eq:GLmin}
\frac{\delta \Omega^{eff}}{\delta \phi({\bf r})}=0.
\end{equation}
For  spatially uniform fields the above minimum condition
yields the following relation between $\phi$ and $|{\bf u}|$:
\begin{equation}
\label{eq:phiurel}
\phi(a_1+2a_{12}|{\bf u}|^2+2b_{12}|{\bf u}|^4)+r_1|{\bf u}|^2+r_2|{\bf u}|^4=0.
\end{equation}
Near a tricritical point both the fields
$\phi$ and $|{\bf u}|$  are small. Therefore it is 
sufficient to consider only  
 a linear coupling between $|{\bf u}|^2$ and $\phi $, neglecting the
 higher order terms: 
\begin{equation}
\label{eq:lincop}
\phi =-\frac{r_1}{a_1}|{\bf u}|^2+2\frac{r_1a_{12}}{a_1^2}|{\bf u}|^4+O(|{\bf u}|^6).
\end{equation}
Inserting   Eq.~(\ref{eq:lincop}) into  Eqs.~(\ref{eq:omegaeff2}) and (\ref{eq:omegaeffin}) we obtain
\begin{equation}
\label{eq:efu}
\frac{\Omega^{eff}}{V}=\frac{1}{2}a_2|{\bf u}|^2+(b-\frac{1}{2}\frac{r_1^2}{a_1})|{\bf u}|^4+e'|{\bf u}|^6,
\end{equation}
with
\begin{equation}
\label{eq:d'}
e'=a_{12}\frac{r_1^2}{a_1^2}-\frac{r_1r_2}{a_1}+e.
\end{equation}
The condition $a_2=0$ yields  the equation for the critical line
which is  in agreement with Eq.~(\ref{eq:critcurve}).
If  $a_1$ is negative,
 $e'$ is positive  and  there is a tricritical point determined by
\begin{equation}
\label{eq:contri}
a_2=0~~~~~~~~~b-\frac{1}{2}\frac{r_1^2}{a_1} =0.
\end{equation}
The solution of these two equations coincides with the expressions
given  by Eqs.~(\ref{eq:tritemp}) and
 (\ref{eq:tritempQ}), i.e., the  tricritical point of this LG model  is located at the same
temperature and  concentration of $^3$He atoms as the tricritical point
in the VBEG model
studied in Sec.~\ref{sec:mf} within mean field approximation.

\section{Summary}
\begin{figure}[h]
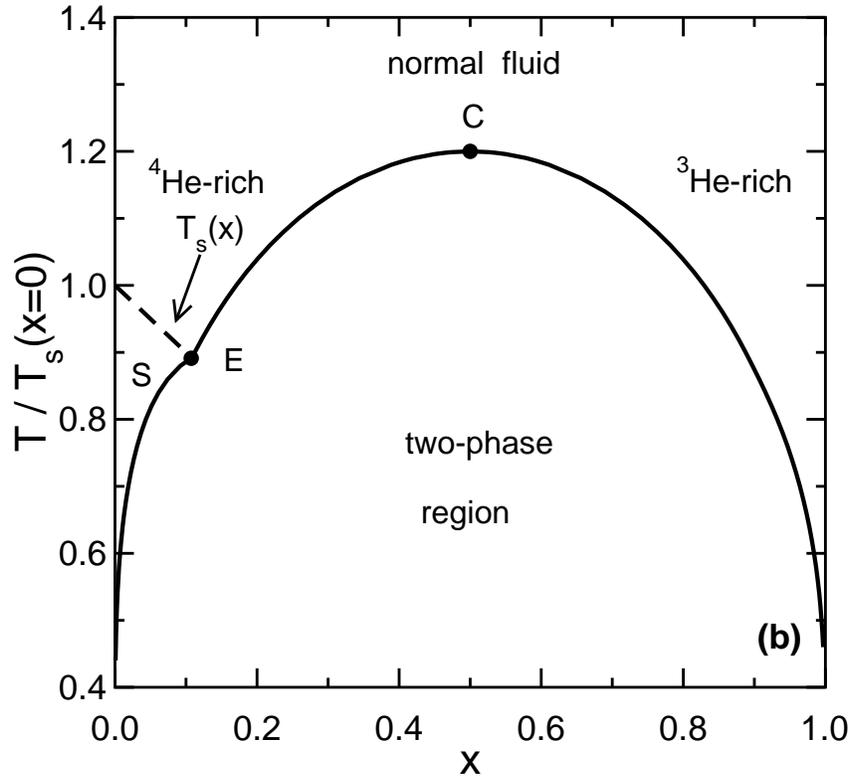

\begin{center}
\includegraphics[scale=0.6]{r3a.eps}
\\[2cm]
\includegraphics[scale=0.6]{r3.eps}
\end{center}
\vskip25pt
\caption{ Phase diagram in the $(\Delta,T)$, (a), and  $(x,T)$, (b), plane
for the model given by  Eq.~(\ref{eq:ham}) 
obtained whithin mean-field theory  for $K/J=2.4$. $x$ is the $^3$He concentration.
 There are three phases which
can be identified as a $^3$He-rich normal fluid ($M=0, x=1-Q$ large), a  $^4$He-rich normal fluid ($M=0, x$ small), and a  $^4$He-rich superfluid  S ($M\ne 0, x$ small).
In (a) the dashed line represents second-order phase transitions and corresponds to the $\lambda$-line; full lines
are the loci of first-order phase transitions.
For this value of $K/J$ there is no tricritical point.
The  $\lambda$-line of second-order phase transitions  terminates at the phase-separation
curve at the critical end point E.
 The two-phase region in (b) and the line of first-order phase transitions in (a) end 
at a critical point C. 
The coordinates of the critical points are:
$C=(\Delta/T_s(0)=2.4, T/T_s(0)=1.2)$, $E=(\Delta/T_s(0)=2.4, T/T_s(0)\approx 0.89)$
 and $C=(T/T_s(0)=1.2, x=1/2)$ and
$E=(T/T_s(0)\approx 0.89, x\approx 0.107)$. In (b) the $\lambda$-line is
given by $T_s(x)/T_s(0)=1-x$ (see Eq.(\ref{eq:critcurve})). The two-phase region
in (b)  between C and E is symmetric about $x=1/2$.
 } 

\label{fig:type1} 
\end{figure}

\begin{figure}[h]
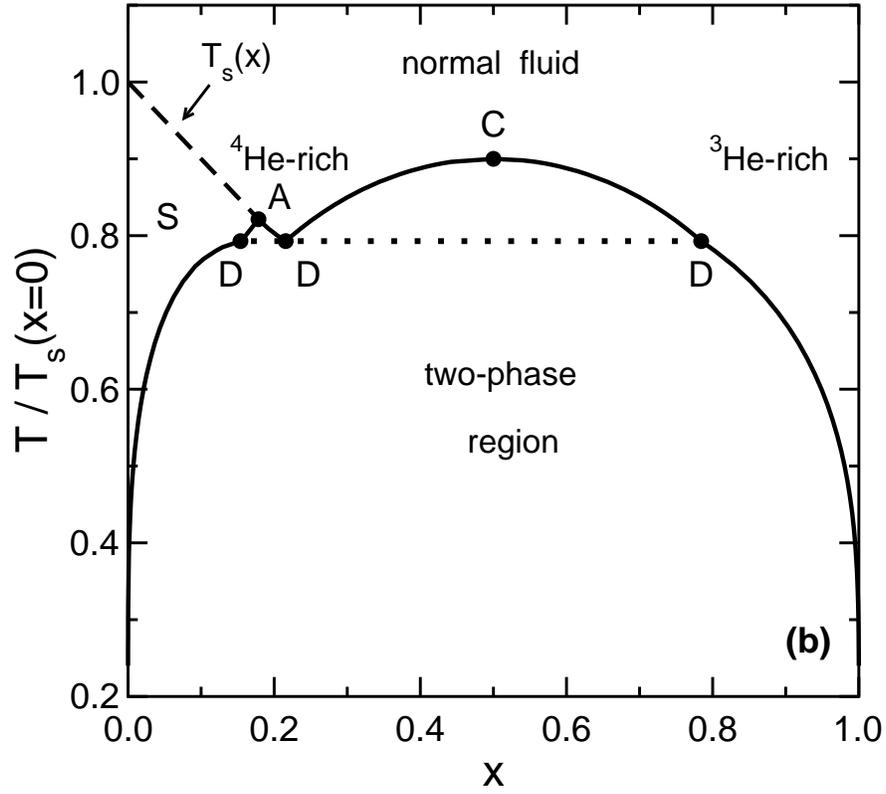
 
\begin{center}
\includegraphics[scale=0.6]{r2a.eps}
\\[2.5cm]
\includegraphics[scale=0.6]{r2.eps}
\end{center}
 \vskip35pt 
\caption{Same as in Fig.1 for $K/J=1.8$. For this value of $K/J$ the
$\lambda$-line ends at a tricritical point A beyond which there is
a first-order phase transition between the $^4$He-rich superfluid S
and the $^4$He-rich normal fluid. At even lower temperatures there is
a triple point D. The coordinates of these points are:
A $=(\Delta/T_s(0)\approx 1.704$, $T/T_s(0)\approx 0.821)$,
C $=(\Delta/T_s(0)=1.8$, $T/T_s(0)=0.9)$, D $=(\Delta/T_s(0)=1.8$,
$T/T_s(0)\approx 0.793)$ and A $=(T/T_s(0)\approx 0.821$, $x\approx 0.179)$,
C $=(T/T_s(0)=0.9, x=1/2)$ and
D $=(T/T_s(0)=0.793, x_1\approx 0.154$, $x_2\approx 0.216, x_3\approx 0.784)$.
In (b) there are two-phase coexistence regions below A and below C which
join for three-phase coexistence at D (dotted curve).}
\label{fig:type2} 
\end{figure}

\begin{figure}[h]
\begin{center}
\includegraphics[scale=0.6]{r1a.eps}
\\[2.5cm]
\includegraphics[scale=0.6]{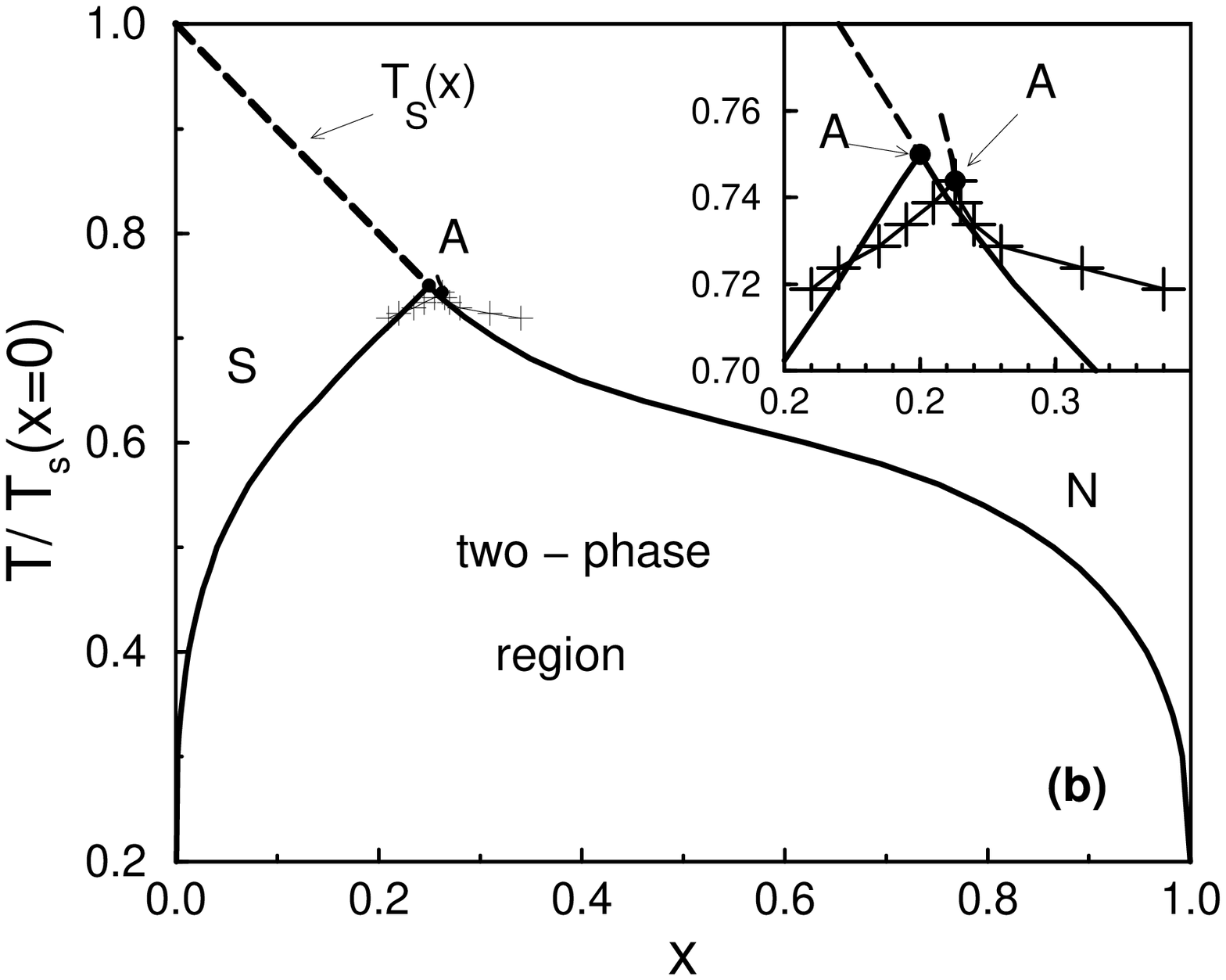}
\end{center}
\vskip15pt
\caption{ 
Same as in Figs. 1 and 2 for $K/J=1$. 
The $\lambda$-line $T_s(x)$ and the
first-order phase separation line meet at the  tricritical point A.
The $^3$He-rich 'normal' fluid phase is denoted by N.
In (b)  Monte-Carlo data for the phase boundaries are indicated by pluses
which are connected by thin lines representing the Monte Carlo phase
boundaries. The inset shows  the results on an expanded scale near
the tricritical point A. The coordinates of the tricritical point
within mean-field theory are: $A=(\Delta/T_s(0)\approx 0.776,
T/T_s(0)= 0.75)$ and $A=( T/T_s(0)= 0.75, x=0.25)$. In (b) the tricritical
point A as obtained from Monte Carlo data is also denoted by a dot and
has the coordinates $(T/T_s(0) = 0.744, x = 0.26)$.
} 
\label{fig:type3} 
\end{figure}

\begin{figure}[h] 
\centerline{\mbox{\epsffile{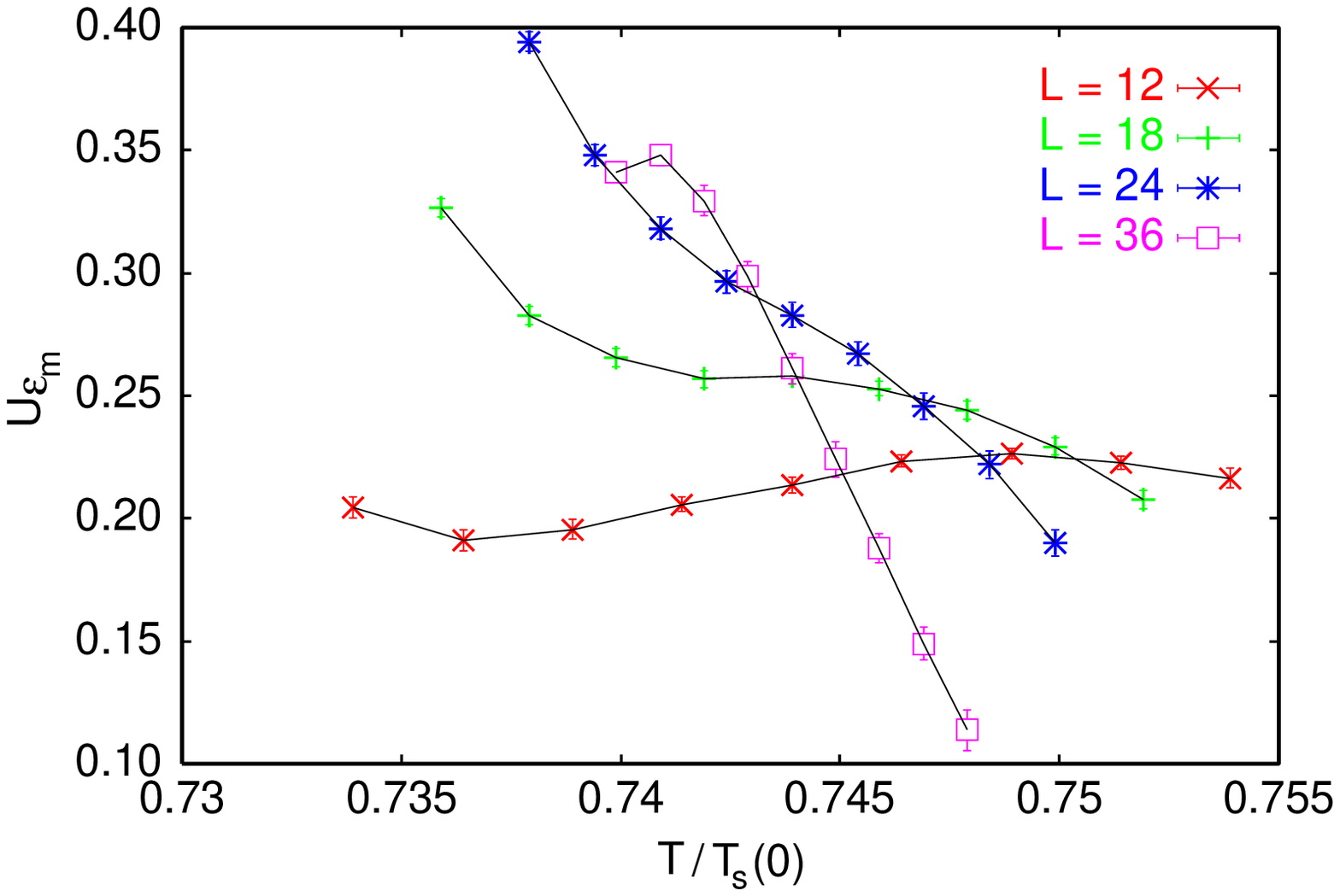}}} 
\vskip15pt
\caption{Cumulant ratio $U_{\varepsilon_m}$ according to Eq.
(\protect\ref{eq:cumEmagn}) as function of temperature along the
straight path used in Fig. \protect\ref{fig:PofnT} for $L = 12$ ($\times$),
$L = 18$ (+), $L = 24$ ($*$), and $L = 36$ ($\Box$). Pairs of symbols are
connected linearly to guide the eye. A unique crossing cannot be identified
(see main text).}
\label{fig:cumuE}
\end{figure}

By using molecular-field approximations and Monte Carlo simulations
we have investigated a three-dimensional version of the generalized spin-1
{\em B}lume-{\em E}mery-{\em G}riffith model (Eq. (\ref{eq:ham})) of
$^3$He-$^4$He mixtures with a two-component vector order parameter, mimicing
the phase of the wavefunction of $^4$He atoms. This  work is a first step to
study the Casimir force and other surface and finite-size effects in
$^3$He-$^4$He mixtures films near their tricritical point. We have obtained
the following main results:
\begin{enumerate}
\item{} The topology of the phase diagram depends on the ratio of the
interaction parameters $K/J$, where $K$ is related to the
$^\alpha$He-$^\beta$He interactions (Eq. (\ref{eq:parK})) and $J$ to the
superfluid  density (Eq. (\ref{eq:paJ})).
There are three different types of the phase diagram, which are
similar to those found in the BEG model within the molecular-field
approximation. For large values of $K/J$, i.e., for  $K/J>2.01681$,
there exist three different phases:
a $^3$He-rich 'normal' fluid, a $^4$He-rich 'normal' fluid,
and a $^4$He-rich superfluid (see Fig.~\ref{fig:type1}).
As the temperature is lowered, the mixed normal fluid phase separates
into two 'normal' fluids  differing by the concentration $x$ of $^3$He.
This phase separation ends at a critical point. At lower temperature, the
phase separation is into a superfluid and a 'normal' fluid. The $\lambda$-line
$T_s(x)$ of second-order phase transitions between a $^4$He-rich
'normal' fluid and a $^4$He-rich superfluid terminates at the $^4$He-rich
branch of the phase-separation curve at the critical end point.
This 'critical end-point' type of the phase diagram was the only one found
in previous studies \cite{cardy:79:0,berker:79:0} of the two-dimensional
version of the model. In three dimensions we find two additional topologies
of the phase diagram as the ratio $K/J$ is decreased. For $1.4298 < K/J<2.01681$ the phase diagram is the richest (see Fig.~\ref{fig:type2}).
As in the previous case, there are three different phases and a critical point
of the phase-separation curve between two 'normal' fluids
differing by the  concentration of $^3$He. In addition, there is a tricritical
point at the end of the $\lambda$-line  beyond which a first-order phase
transition between a $^4$He-rich superfluid and  a $^4$He-rich 
'normal' fluid takes place. There is also a triple point at which 
three different  phases coexist at different concentrations.
For $K/J < 1.4298$ the phase diagram  simplifies (see Fig.~\ref{fig:type3}).
There is no longer a $^4$He-rich 'normal' fluid phase and a critical point.
The $\lambda$-line meets the first-order phase separation line between 
$^4$He-rich superfluid and a $^3$He-rich 'normal' fluid at the tricritical
point. The $\lambda$-line is given by $T_s(x)=zJ(1-x)/2$, where $z$ is the
coordination number of the lattice. The temperature of the tricritical point
is $T_A/ T_s(0)=(1+2K/J)/(2+2K/J)$ (Eq.~(\ref{eq:tritemp})) and the
concentration $x_A$ of $^3$He at this point is given by $T_A/ T_s(0)=1-x_A$
(Eq.~(\ref{eq:tritempQ})). This type of the phase diagram is similar to that
observed experimentally, although in our model $x_A$ is  always smaller than
0.5 whereas $x_A^{exp}=0.669$.

\item{} The existence of the tricritical point is confirmed by Monte Carlo
simulations and in the units $(T_t/T_s(0), \Delta_t/J)$ it coincides with the
mean-field prediction remarkably well (see Fig. \ref{fig:type3}). At the
tricritical point the order parameter distribution function takes its
mean-field form, where the presence of logarithmic corrections could not be
excluded within the accuracy of the existing data. On the other hand
finite-size scaling of several thermodynamic
quantities reveals the presence of logarithmic corrections in
accordance with theoretical expectations. The two-phase coexistence line in
the $(T,\Delta)$ plane of the phase diagram has been determined form a
constant weight ratio criterion for energy and density histograms. The
location of the tricritical point is also indicated by a crossing of the
cumulant ratio of the magnetic portion of the energy density measured
along the coexistence line (see Fig. \ref{fig:cumuE}). We conclude that
mean-field theory provides a reliable approach for studying the VBEG model
in $d=3$.

\item{} Starting from the VBEG model and discretizing the orientations of
the spin vector we have derived the continuum Landau-Ginzburg model for
$^3$He-$^4$He mixtures near the tricritical point encompassing the
concentration field $\phi$ and a two-component vector field ${\bf u}$
corresponding to the orientational order. In the effective Hamiltonian we
consider the modulus of ${\bf u}$ up to its sixth power and the field $\phi$
up to quadratic terms, which is sufficient to study a tricritical point.
The coupling constants appearing in this Landau-Ginzburg theory are given
explicitly in terms of thermodynamical quantities, the temperature, the
mean concentration of $^4$He, and the two interaction parameters J and K
characterizing the VBEG model. Mean-field theory for this LG model yields
the same equation for the critical $\lambda$-line as the molecular-field
approximation for the lattice VBEG model. The LG model provides a linear
coupling between $|{\bf u}|^2$ and $\phi$ which yields the same coordinates
of the tricritical point as the lattice VBEG model.

\end{enumerate}

\acknowledgments
A.M. is grateful for the hospitality accorded by the
Max-Planck-Institut f{\"u}r Metallforschung in Stuttgart, Germany. She
appreciates fruitful discussions with Alina Ciach.
This work was partially funded by KBN grant No.4 T09A 066 22.

\hfill\eject

\hfill\eject

\begin{thebibliography}{99}

\bibitem{krech:99:0} M. Krech, {\it The Casimir Effect in  Critical System}   
(World Scientific, Singapore, 1994), and references therein; J. Phys. Condens. Matter {\bf 11}, R391 (1999).

\bibitem{garcia:99:0} R. Garcia and M. H. W. Chan, Phys. Rev. Lett. {\bf 83}, 1187 (1999).

\bibitem{krech:91:0} M. Krech and S. Dietrich, Phys. Rev. Lett. {\bf 66}, 345 (1991); {\bf 67}, 1055 (1991).
\bibitem{krech:92:0} M. Krech and S. Dietrich, Phys. Rev. A {\bf 46}, 1886 (1992).

\bibitem{krech:92:1}  M. Krech and S. Dietrich, Phys. Rev. A {\bf 46}, 1922 (1992).

\bibitem{law:99:0}A. Mukhopadhyay and Bruce M. Law, Phys.~Rev.~Lett. {\bf 83}, 772 (1999);
 a quantitative understanding of these experimental data has not yet been reached. 

\bibitem{garcia:02:0} R. Garcia and M. H. W. Chan, Phys. Rev. Lett. {\bf 88}, 086101 (2002).

\bibitem{balibar:02:0} T. Ueno, S. Balibar, T. Mizusaki, F. Caupin, and
E. Rolley, Phys. Rev. Lett. {\bf 90}, 116102 (2003); T. Ueno, S. Balibar, T.
Mizusaki, F. Caupin, M. Fechner, and E. Rolley, J. Low Temp. Phys. {\bf 130},
543 (2003).

\bibitem{laheurte:77:0} J.-P. Laheurte, J.-P. Romagnan, and W. F. Saam, Phys. Rev. B {\bf 15}, 4214 (1977).

\bibitem{leibler:84:0} S. Leibler and L. Peliti, Phys. Rev. B {\bf 29}, 1253 (1984); L. Peliti and S. Leibler, J. Physique Lett. {\bf 45}, L591 (1984).

\bibitem{macqueeney:84:0} D. McQueeney, G. Agnolet, J. D. Reppy, Phys. Rev. Lett. {\bf 52}, 1325 (1984).

\bibitem{kosterlitz:80:0}  J. M. Kosterlitz and D. J. Thouless, J. Phys. C {\bf 5}, L124 (1972).

\bibitem{dietrich:91:0} S. Dietrich, in {\it Phase Transitions and Critical Phenomena}, 
edited by C.~Domb and J.~L.~Lebowitz (Academic, London, 1988), vol.~12, p.~1.

\bibitem{graf:67:0} G. Ahlers and D. S. Greywall, Phys. Rev. Lett. {\bf 29}, 849 (1972); H. A. Kierstead, J. Low Temp. Phys. {\bf 35}, 25 (1979); E. H. Graf, D. M. Lee, and J. D. Reppy, Phys. Rev. Lett. {\bf 19}, 417 (1967).

\bibitem{cardy:79:0} J. L. Cardy and D. J. Scalpino, Phys. Rev. B {\bf  19}, 1428 (1979).

\bibitem{berker:79:0}  A. N. Berker and  D. R. Nelson, Phys. Rev. B {\bf 19},
2488 (1979). 

\bibitem{blume:71:0}M. Blume, V. J. Emery, and R. B. Griffiths, Phys. Rev. A {\bf 4}, 1071 (1971).

\bibitem{bell:89:0} see for example G. M. Bell and D. A. Lavis, {\it Statistical Mechanics
 of Lattice Models}, series in "Mathematics and its Applications"  (Ellis Horwood Ltd, Chichester, 1989).



\bibitem{bishop:78:0} D. J. Bishop and J. D. Reppy, Phys. Rev. Lett. {\bf 40}, 1727 (1978).


\bibitem{mermin:66:0}  N. D. Mermin and  H. Wagner, Phys. Rev. Lett. {\bf 17}, 1133 (1966).


\bibitem{book} see for example P. M. Chaikin and T. C. Lubensky, {\it Principles of
Condensed Matter Physics} (Cambridge University Press, 1995).

\bibitem{comment} In the other approach to determine variational minima to Eq. (\ref{eq:varfe}) a parametrization of $\rho _i$ in terms of  order parameters
$<\phi _i>$ ( for example $Q$, $M_x$, and $M_y$) can be chosen. This parametrization must satisfy the constraints
$Tr \rho _i=1$ and $Tr \rho _i\phi _i = <\phi _i>$.
 The variational parameters are simply order parameters  and $F_{\rho_0}$ is the Helmholtz free energy functional $F(<\phi>)$. This procedure is less general than the one used in the present paper.

\bibitem{abramowitz} {\it Handbook of Mathematical Functions}, edited by 
Abramowitz M., and Stegun I. A., (Dover Publications Inc., New York, 1972).


\bibitem{fisher:91:0} M. E. Fisher and M. C. Barbosa, Phys Rev. B {\bf 43}, 11177 (1991).

\bibitem{Metropolis}
N. Metropolis, A. W. Rosenbluth, M. N. Rosenbluth, A. H. Teller,
and E. Teller, J. Chem. Phys. {\bf 21}, 1087 (1953).

\bibitem{Wolff89}
U. Wolff, Phys. Rev. Lett. {\bf 62}, 361 (1989).

\bibitem{CFL93}
K. Chen, A. M. Ferrenberg, and D. P. Landau, Phys. Rev. B {\bf 48}, 3249
(1993).

\bibitem{Hybrid}
J. A. Plascak, A. M. Ferrenberg, and D. P. Landau, Phys. Rev. E {\bf 65},
066702 (2002).

\bibitem{cluerr}
A. M. Ferrenberg, D. P. Landau, and Y. J. Wong, Phys. Rev. Lett. {\bf
69}, 3382 (1993); L. N. Shchur and H. W. J. Bl\"ote, Phys. Rev. E {\bf
55}, R4905 (1997).

\bibitem{SimTemp}
T. Nagasima, Y. Sugita, A. Mitsutake, and Y. Okamoto, Comp. Phys. Commun.
{\bf 146}, 69 (2002).

\bibitem{Histogram}
A. M. Ferrenberg and R. H. Swendsen, Phys. Rev. Lett. {\bf 61}, 2635 (1988);
{\bf 63}, 1195 (1989).

\bibitem{Nigel}
N. B. Wilding and P. Nielaba, Phys. Rev. E {\bf 53}, 926 (1996).

\bibitem{TcXY}
M. Krech, unpublished, see also 
W. Janke and H. Kleinert, Nucl. Phys. {\bf B270}, 135 (1986)

\bibitem{LawSar}
D. Lawrie and S. Sarbach, in {\em Phase Transitions and Critical
Phenomena}, edited by C. Domb and J. L. Lebowitz (Academic, London,
1984), vol. 9, p.2.

\bibitem{ciach:96:0}  A. Ciach, J. Chem. Phys. {\bf 104}, 2376 (1996).


\end{thebibliography}
\end{document}